\documentclass[preprint]{aastex}        %finished
\usepackage{graphicx,times,natbib,lscape}

\newcommand{\sersic}{{S\'{e}rsic }}
\def\arcsec{$^{\prime\prime}$}

\def\spose#1{\hbox to 0pt{#1\hss}}
\def\lta{\mathrel{\spose{\lower 3pt\hbox{$\mathchar"218$}}
     \raise 2.0pt\hbox{$\mathchar"13C$}}}

\shorttitle{Color Gradient of $z\sim2$ early-types in WFC3/GOODS}
\shortauthors{Guo et al.}

\begin{document}

\title{Color and Stellar Population Gradients in Passively Evolving Galaxies at $z\sim2$ from HST/WFC3 Deep Imaging in the Hubble Ultra Deep Field}
\author{Yicheng Guo$^{1,2}$, Mauro Giavalisco$^{1}$, Paolo Cassata$^{1}$, 
Henry C. Ferguson$^{3}$, Mark Dickinson$^{4}$, Alvio Renzini$^{5}$, 
Anton Koekemoer$^{3}$, Norman A. Grogin$^{3}$, Casey Papovich$^{6}$, 
Elena Tundo$^{7}$, Adriano Fontana$^{8}$, 
Jennifer M. Lotz$^{3}$, Sara Salimbeni$^{1}$}
\affil{$^1$ Astronomy Department, University of Massachusetts,
710 N. Pleasant Street, Amherst, MA 01003, U.S.A.}
\affil{$^2$ email: {\texttt yicheng@astro.umass.edu}}
\affil{$^3$ Space Telescope Science Institute, 3700 San Martin Drive, 
Baltimore, MD, 21218, U.S.A.} 
\affil{$^4$ NOAO-Tucson, 950 North Cherry Avenue, Tucson, AZ 85719, U.S.A.}
\affil{$^5$ INAF - Osservatorio Astronomico di Padova, Vicolo
dell'Osservatorio 5, I-35122, Padova, Italy}
\affil{$^6$ George P. and Cynthia Woods Mitchell Institute for Fundamental 
Physics and Astronomy, Department of Physics, Texas
A\&M University, College Station, TX 77843-4242, U.S.A.}
\affil{$^7$ INAF - Osservatorio Astronomico di Trieste, Via Tiepolo 11, I-34131 
Trieste, Italy}
\affil{$^8$ INAF - Osservatorio Astronomico di Roma, via Frascati 33,
Monteporzio-Catone (Roma), I-00040, Italy}

%\date{{\sc Version 6a: } \today }

\begin{abstract}
We report the detection of color gradients in six massive (stellar mass ${\rm
  (M_{star}) > 10^{10}}$ M$_{\odot}$) and passively evolving (specific star
formation rate (SSFR) $ < {\rm 10^{-11} yr^{-1}}$) galaxies at redshift
$1.3<z<2.5$ identified in the Hubble Ultra Deep Field (HUDF)
using ultra--deep {\it HST} ACS and WFC3/IR images.
%, the deepest optical and
% near-IR observations obtained to date.
After carefully matching the different PSFs, we obtain color maps and
multi--band optical/near--IR photometry (BVizYJH) in concentric annuli, 
% chosen
% to optimally sample the color gradients across the light profile of each
% galaxy, 
from the smallest resolved radial distance ($\approx 1.7$ kpc) up to
several times the H--band effective radius. We find that the inner regions of
these galaxies have redder rest-frame UV--optical colors (U-V, U-B and B-V)
than the outer parts. The slopes of the color gradient have no obvious
dependence on the redshift and on the stellar mass of the galaxies. They do
mildly depend, however, on the overall dust obscuration (E(B-V)) and
rest--frame (U-V) color, with more obscured or redder galaxies having steeper
color gradients. The $z\sim 2$ color gradients are also steeper than those of
local early--type ones.
The gradient of a single parameter (age, extinction or metallicity) cannot
fully explain the observed color gradients. Fitting the spatially resolved
{\it HST} seven--band photometry to stellar population synthesis models, we
find that, regardless of assumptions on the metallicity gradient, the redder
inner regions of the galaxies have slightly higher dust obscuration than the
bluer outer regions, implying that dust partly contributes to the observed
color gradients, although the magnitude depends on the assumed extinction law.
% (the strongest gradient, ${\rm \Delta E(B-V) / \Delta log(R) \sim -0.07}$,is
% found using the Calzetti extinction curve). 
Due to the age--metallicity degeneracy, the derived age gradient depends on
the assumptions for the metallicity gradient. We discuss the implications of 
a number of assumptions for metallicity gradients on the formation and 
evolution of these galaxies. 
% : 1) a flat metallicity gradient
% implies that the outer regions of the galaxies are younger than the inner
% regions, with age gradient $\alpha_t = {\rm \Delta log(t) / \Delta log(R) \sim
%   -0.1}$; 2) a metallicity gradient like the one of local early--type galaxies
% implies that the stellar populations in the outer regions have similar age as
% those in the central regions; and 3) the metallicity gradients predicted by
% the monolithic collapse implies that the outer regions are older than those in
% the center ($\alpha_t \sim 0.15$). 
% While the mass--size relationship of these
% galaxies undergoes a dramatic evolution from $z\sim2$ to the present, this is
We find that the evolution of the mass--size relationship from $z\sim2$ 
to the present cannot be driven by in--situ extended star formation, which
implies that accretion or merger is mostly responsible for the growth of
their stellar mass and size. 
% A flat or local metallicity gradient are consistent with this scenario,
% but inconsistent with major merger as the primary driver of the
% evolution. 
The lack of a correlation between the strength of the color gradient and 
the stellar mass 
% despite the latter being the most robust
% parameter from broad--band spectral energy distribution (SED) fitting, 
argues against the metallicity gradient predicted by the monolithic collapse
scenario, which would require significant major mergers to evolve into the 
one observed at the present.
\end{abstract}


\section{Introduction}
\label{intro}

The stellar mass (${\rm M_{star}}$) of ``spheroids'', namely elliptical
galaxies and bulges of spiral galaxies, mostly consists of old stars that
formed at high redshift, e.g. $z>2$ (Renzini 2006). The spheroids segregate
$\approx 60$\% of all the stars in the local universe
\citep{hogg02,bell03,driver06a}, and thus the mechanisms that led to their
assembly are key to the evolution of galaxies in general. But while there is
agreement on the age of the stars of the spheroids, how these stars got
together and formed the body of ellipticals and bulges remains an open issue.

During the past several years, galaxies at $z>1$ with ${\rm M_{star}}$ and SED
similar to those of local early-type galaxies have been identified and studied
in relatively large numbers thanks to the increased availability of deep
optical and near--IR photometry from large--area surveys
\citep[e.g.,][]{thompson99, franx03, daddi04bzk} as well as spectral data
\citep[e.g.,][]{kriek06a, kriek06b, cimatti08, onodera10p}. More recently, an
increasing number of studies seem to show that the number density of massive
galaxies with very low specific star formation rate (SSFR) undergoes rapid
evolution between $z\sim2$ and $z\sim1$ \citep[e.g.,][Cassata in preparation]{fontana06, arnouts07,
  Ilbert10}. The physical mechanisms responsible for this
apparently rapid assembly of passively evolving massive galaxies remain
unknown. Equally unknown is if this is just the assembly of the stellar bodies
of the massive galaxies, or if the fraction of stars locked in passively
evolving systems is also evolving accordingly. Various formation mechanisms,
for example, merger \citep[e.g.,][]{brinchmann00, lefevre00, delucia06,
  naab07, naab09b} and feedback \citep[e.g.,][]{benson03, croton06} have been
proposed to explain the rapid emergence of massive passively evolving galaxies 
(PEGs) during this redshift
range, which is also the cosmic epoch when star formation in the universe is
at its peak. Such mechanisms would leave distinguishable imprints on the color
and stellar population gradients of massive PEGs.  For example, a major merger
(i.e. mass ratio approximately unitary) of gas--rich galaxies would form a
spheroid and trigger a bursts of central star-formation, which would leave a
blue core to the spheroid \citep{menanteau01a, menanteau04, daddi05}. Or, if
massive PEGs mostly assemble their masses through dry mergers or mergers that
do not induce central star-formation, they would generally not have blue
cores. Thus, studying color gradients and their implications on the stellar
population gradients of massive PEGs is expected to provide important clues on
the formation of massive PEGs at $z\sim2$.

Related to the formation mechanisms is the issue of the subsequent evolution
of the massive PEGs. Recent work \citep[e.g.,][]{daddi05, trujillo06,
  trujillo07, vandokkum08, cassata10} shows that many massive PEGs at $z>1.5$
are, on average, $\sim$5 times smaller and $\sim50$ times denser than their
local counterparts with similar mass. The physical mechanisms proposed to
explain this apparently dramatic evolution of size include major merger
\citep[e.g.,][]{hopkins09c,vanderwel09}, minor merger
\citep[e.g.,][]{naab09a}, adiabatic expansion \citep[e.g.,][]{fan08}, and
mass-to-light gradients \citep[e.g.,][]{hopkins10a}. Others
\citep[e.g.,][]{hopkins09b, mancini10} have suggested that the small size of
some PEGs at high redshift may be due to an observational bias such that the
low surface-brightness halos surrounding these PEGs are not detected by
current near-IR observations; if these missing halos were detected, the
derived size of high-z massive PEGs would be similar to that of their local
counterparts. In order to answer the question whether the observed strong size
evolution of massive PEGs from $z\sim2$ to $z\sim0$ is physical or not,
near-IR observations with high sensitivity are required to measure the color
and stellar population distributions of massive PEGs to large radius.

Color gradients in early type galaxies have been known for about thirty years
\citep{faber72} and widely studied in local galaxies
\citep[e.g.,][]{peletier90a, tamura00, labarbera05, labarbera09, gonzalez11}, but no
information is currently available on color gradients in PEGs at high redshift
($z\sim2$), because of instrumental limitations on sensitivity and angular
resolution, given the compact size of such sources, and the lack of spectral
coverage of the rest--frame optical SED. Ground-based observations suffer
from poor resolution and/or wavelength--dependent and unstable Strehl
ratio. Sensitivity to low--surface brightness regions is also limited due to
the high and variable sky background. For example, the typical full-width
half-maximum (FWHM) of the point spread function (PSF) of VLT ISAAC Ks--band
images is about 0.5\arcsec, corresponding to $\sim$4 kpc for a galaxy at
$z\sim2$. This size is almost 4 times of the average effective radius of a
PEGs with ${\rm M_{star}}$ = ${\rm 10^{10}M_\odot}$ at $z\sim2$ \citep[][and
  reference therein]{cassata10}. Even if upcoming adaptive optics systems
reach near--{\it HST} resolution in the K band, performance degrade rapidly at
shorter wavelength so that making robust color maps is not yet feasible. To
sample the color and stellar population gradients of PEGs at $z\sim2$ at the 
$\sim$kpc scale, a minimum angular resolution of about $\sim0.1$\arcsec\ is 
required at both optical and near--IR wavelength.
%to sample color gradients at the $\sim$kpc scale. 
Although {\it HST} NICMOS-1 and NICMOS-2 have such required resolution,
their small fields of view and low throughput make them inconvenient for
surveying large sky area and observing distant and faint galaxies. The
detailed study on color gradients of a large sample of high-redshift
early-type galaxies is only now available thanks to the WFC3/IR imager
on-board of {\it HST}.

In this work we use the HUDF {\it HST}/ACS images in combination with recent
WFC3 near--IR deep images in the same field to measure the color gradients of
a sample of massive PEGs at $z\sim2$ to about 10 times their effective radius
and inferred corresponding gradients of physical properties of the stellar
populations. We measure color gradients for these galaxies in a series of
concentric annuli from the ACS and WFC3 images and fit the spatially resolved
SED to stellar population synthesis models to derive the corresponding
gradients of stellar population parameters (SSFR, age and extinction), looking
for trends between the color gradient characteristics and the stellar
population properties in an attempt to derive clues on their origins.

% THERE IS NO INFORMATION CONTENT IN THIS PARAGRAPH. ELIMINATE
%The paper is organized as follow. The data sets used in the paper are
%described in \S\ref{data}. In \S\ref{pegs}, we present our sample of massive
%PEGs and their photometry of concentric rings.  In \S\ref{cg}, we study the
%observed color gradients and their relations with the photometrically-measured
%physical properties of massive PEGs.  In \S\ref{spg}, we study the stellar
%population gradients of massive PEGs. We focus on the age, SSFR, and stellar
%density gradients in this paper.  The summary and conclusions are presented in
%\S\ref{summary}. 
Throughout we adopt a flat ${\rm \Lambda CDM}$ cosmology with
$\Omega_m=0.3$, $\Omega_{\Lambda}=0.7$ and use the Hubble constant in terms of
$h\equiv H_0/100 {\rm km~s^{-1}~Mpc^{-1}} = 0.70$.  All magnitudes in the
paper are in AB scale \citep{oke74} unless otherwise noted.

\section{The Data}
\label{data}

In addition to the {\it HST }/ACS and WFC3/IR images in the HUDF, the data
used in this paper also include panchromatic multi--wavelength photometry
obtained as part of the GOODS program, as the HUDF field is embedded in the
GOODS south field. The long wavelength baseline of the GOODS photometry
enables us to reliably select PEGs based on photometrically--derived stellar
mass and SSFR, while the deep {\it HST} optical and NIR images allow us to
obtain color maps of the galaxies with a resolution of $\sim 1$ kpc.
% This consists of ground--based ultradeep
% U \citep{nonino09} and JHK \citep{giavalisco04} images and Spitzer ISAAC and
% MIPS 24 $\mu$ images, from which matched photometry from U to Irac 4 band,
% also including all
% the {\it HST} bands, has been obtained using an object template--fitting algorithm dubbed
% TFIT \citep[the photometric catalog goes under the name of GUTFIT, see][and Grogin et al. in prep.]{laidler07},
% % the photometric catalog goes under the name of GUTFIT Laidler et al. 2007; 
% % Grogin et al. in prep.) 
% which iteratively deblends
% confused images. The algorithm, which requires positional priors (typically
% from either ACS or WFC3/IR detections) and accurate measures of the PSF for
% each image, has been extensively tested with simulations and shown to yield
% accurate photometry. The {\it HST} data include the ultra--deep optical ACS
% observations of HUDF and the first epoch of similarly deep near--IR WFC3/IR
% observations acquired during the HST Cycle--17 General Observer Program by
% Illingworth et al.  \citep[HSTGO-11563, see]{bouwens10}. 

The GOODS south field has been observed with various telescopes and instrument
combinations, from the X--ray to the sub--millimeter and radio. Relevant to our
analysis here is the VLT/VIMOS ultra--deep U--band imaging \citep[]{nonino09},
as well as {\it HST}/ACS BViz \citep{giavalisco04}), VLT/ISAAC JHK,
Spitzer/IRAC 3.6, 4.5, 5.7, 8.0 $\mu$m, and Spitzer/MIPS 24 $\mu$m imaging.  
Since the resolution of
images significantly changes from optical- to IR-band, we use an object
template-fitting software dubbed TFIT \citep{laidler07} to obtain matched multi--band
photometry. TFIT requires position priors and light profile templates drawn
from a high-resolution image (the ACS z-band image in our work) and accurate
measures of the PSF for all images with various resolutions.
% have created source catalogs using TFIT to TFIT requires the knowledge of
% the PSF for each image to be analyzed, and uses the spatial position and
% light profile of sources detected in a high-resolution image (the ACS
% z--band image in this work) as a template.  
It fits the template of an object, whose resolution is now downgraded to that
of low-resolution images, to the images of the object in low-resolution bands,
with the flux in each band left as a free parameter. The best-fit flux in each
band is used as the flux of the object in the band. TFIT can simultaneously
fit several objects that are close enough in the sky so that the deblendding
effect of these objects on the flux measurement would be minimized.
Experiments on both simulated and real images show that TFIT is able to
measure accurate isophotal photometry of objects to the limiting sensitivity
of images. The TFIT measured fluxes of bands with resolution lower than the
z-band resolution, together with the SExtractor measured AUTO flux of BViz
bands, are merged to build the GUTFIT catalog (Grogin et al. in prep.).

The ultra--deep ACS images in the HUDF \citep{beckwith06hudf} cover an area
roughly equal to the footprint of the ACS/WFC FOV in the same four filters as
the GOODS ACS program, namely F435W (B), F606W (V), F775W (i), and F850LP (z)
down to a depth of 29.4, 29.8, 29.7, and 29.0 mag (5$\sigma$,
0.35\arcsec-diameter aperture), respectively. We use the publicly available
images, which have been rebinned to the same pixels scale as the GOODS/ACS
mosaic, namely 0.03\arcsec\ /pixel ($0.6 \times$ the original ACS pixel
scale).  

The WFC3/IR data are from the {\it HST} Cycle 17 program GO-11563 (PI:
G. Illingworth), which aims at complementing the HUDF and the two HUDF05
parallel fields \citep{oesch07} with WFC3/IR images in Y (F105W), J (F125W),
and H (F160W) of matching sensitivity, $\sim$29 mag \citep{bouwens10,oesch10}.  Here we use only the
first epoch of the images, released in September 2009, which includes 18 orbits
in Y, 16 orbits in J, and 28 orbits in H. We have carried out our independent
reduction of the raw data, and after rejecting images affected by persistence
in the J band, our final stacks reach 1$\sigma$ surface brightness
fluctuations of 27.2, 26.6 and 26.3 AB/\arcsec\ $^2$ in the three bands,
respectively, over an area roughly equal to the
footprint of the WFC3/IR camera (2.1 \arcsec\ $\times$ 2.1 \arcsec). We have
drizzled the WFC3 images from their original pixel size of 0.121\arcsec
$\times$ 0.135\arcsec\ to 0.03\arcsec\ per pixel to match the scale of the
GOODS and HUDF ACS images.

\section{Passively Evolving Galaxies at $z\sim2$ in HUDF}
\label{pegs}

We select passively evolving galaxies based on their SSFRs
estimated by fitting the GOODS GUTFIT 12--band photometry to
stellar population synthesis models. During the fit, the value of the redshift
parameter is set to either the spectroscopic redshift, when available, or
the photometric redshift, which we have separately measured using the PEGASE
2.0 \citep{pegase} templates. 
%found to provide more accurate results than o5ther templates \citep{Grazian06}. 
Another set of photometric redshifts is
also available in the GOODS--S field\citep{dahlen10}, and we find that
results from both sets are in excellent agreement; both sets achieve
$\sigma(\Delta z/(1+z)) \sim 0.04$ in the redshift range considered here. For
the SED fitting we use the stellar population synthesis models of Charlot \&
Burzual 2009 (CB09), with Salpeter IMF and lower and upper mass limits of 0.1
and 100 ${\rm M_\odot}$, respectively. We also use the $e^{-t/\tau}$--model 
($\tau$--model) to parametrize the star formation history of the galaxies. The free parameters
that are found by the fitting procedure are the stellar mass ${\rm M_{star}}$,
the dust reddening E(B-V), the $\tau$ parameter
and the age $t$ of the stellar populations. We use the Calzetti Law
\citep{calzetti94, calzetti00} to model the obscuration by dust and the
prescription of \citet{madau95} to account for the cosmic opacity by
HI. Finally, we average the best--fit model star--formation history over the
last 100 Myr to derive the current SFR of the galaxy. We estimate the
$1-\sigma$ error bars of the parameters of the best--fit model from
Monte--Carlo simulations, where we perturb the photometric measures using
Gaussian variates with variance set equal to the photometric errors and
re--run the fitting procedure 200 times.

We define passively evolving galaxies in the redshift range $1.3<z<3.0$ as
those whose specific star--formation rate satisfies the relation
\begin{equation}
{\rm SSFR = {SFR \over M_{star}} \le 10^{-11} \hbox{~yr$^{-1}$}},
\label{eq:ssfr}
\end{equation}
and restrict our samples to only include massive systems, namely those with
$M_{star}>10^{10}$ M$_{\odot}$. Among 53 galaxies with 
$M_{star}>10^{10}$ M$_{\odot}$ and $1.3<z<3.0$ in the HUDF, 11 galaxies 
have SSFR${\rm \le 10^{-11} yr^{-1}}$.
We exclude two of them from our sample, because we are
interested in studying color gradients of early--type galaxies, while these
systems have irregular morphology and not well-defined centers, possibly
implying ongoing merging events. We exclude another two galaxies, 
because they have extremely faint NIR fluxes, in fact they have negative J, H
and IRAC fluxes, which result in large uncertainties in their
photometrically-derived physical properties. Finally, an additional galaxy
with an obvious spiral morphology has also been eliminated from the
sample. Although the best--fit SSFR of this galaxy is $10^{-11.11}$ yr$^{-1}$,
the probability distribution function (PDF) of this SSFR measure has two peaks
with similar probability density, one around $10^{-11.11}$ yr$^{-1}$ and the
other $10^{-10.23}$ yr$^{-1}$, implying a substantial probability for this
source to be a star-forming galaxy.

After these exclusions, the final sample includes 6 galaxies, whose GOODS ID,
coordinates, redshift, SED--fitting parameters and H--band effective radius,
measured by \citet{cassata10}, are shown in Table \ref{tb:ppp}.  Five out of
the six galaxies in the sample satisfy pBzK color--color criterion for
passively evolving galaxies at $1.4<z<2.3$ \citep{daddi04bzk} or the analog
VJL criterion for redshift $2<z<3$ (VJL, Guo et~al. in preparation). The last
galaxy, 24626, resides just below the pBzK selection window in the
(B-z) vs.(z-K) color--color plane, most likely because its spectroscopic
redshift, $z=1.31$, is outside of the targeted range of the pBzK
criterion. Four galaxies have spectroscopic redshift, 22704 and 23555 from
\citet{cimatti08}, 24279 from \citet{daddi05} and 24626 from
\citet{vanzella08}. 

Galaxy 23495 has a counterpart in the Chandra Deep Field
South 2--Megasecond catalog \citep{luo08}. It has X-ray luminosity $3.8\times
10^{43} {\rm erg/s}$ and $5.6\times 10^{43} {\rm erg/s}$ in the soft and 
hard band, respectively.  It is not, however, detected in the VLA map by
\citet{kellermann08} and \citet{miller08}. Galaxy 24626 also has a counterpart
in the catalog of \citet{luo08}, but it only has a marginal (S/N $\sim$ 1.3) 
detection in the soft band with X-ray luminosity 
$2.7\times 10^{41} {\rm erg/s}$ and none in the hard band. We re-investigate
the two sources with the newly released Chandra Deep Field South 4--Megasecond
image\footnote{http://cxc.harvard.edu/cda/Contrib/CDFS.html} and find 
similar results. Other four galaxies have no detection in both bands, 
either individually or stacked, in the 4--Megasecond image. Finally, 
all our sample galaxies have no detection at 24 $\mu$m down to a 1$\sigma$ 
limit of 5 $\mu$Jy, consistent with predictions for passively evolving 
galaxies at z $\sim$2 \citep{fontana09}.

Figure \ref{fig:mtg} shows the images of the sample galaxies in the z-- and 
H--bands, as well as their (z-H) color composites. The z--band images have their
original resolution ($\sim$0.12 \arcsec\ ), while the resolution of the (z-H)
color images is that of the H--band images, after PSF matching (see
later). Both the z--band and H--band images show that all the sample galaxies
have spheroidal, early--type morphology, while (z-H) color maps reveal both
analogies and differences among them. All galaxies have a red center and blue
outskirts. Galaxy 24626 has the most well-defined red center and the clearest
color gradient. Galaxy 23555 and 24279 also have well--defined red centers,
but their outskirts are observed at relatively lower S/Ns and the resulting
color gradient is not as clear as that of Galaxy 24626. In the remaining three
galaxies, the location of the red stellar populations is slightly off--center,
with the distance between the centroid and the red center comparable to the
H--band half--light radius of the galaxy. After re--sampling both 
z--band and H--band images to smaller pixel scale (0.01\arcsec\ /pixel) and 
re--registering images, we still find such off--center red cores. Therefore, 
we rule out the sub-pixel image registration issue as the reason of the 
off--center cores. Instead, we suspect the asymmetry in the cores of our 
empirical PSFs could be the reason. However, the use of annuli photometry
with size of a few FWHM largely reduces the influence of the asymmetry of 
PSFs so that it would not impact our results, as our later test in 
\S\ref{sub:psfmatch} shows.
Since we aim at measuring the color
gradients up to $\sim$10 times of the H--band half--light radius, it is still
reasonable to consider that these galaxies, too, have red centers.

\begin{table}[h]
\begin{minipage}{\textwidth}\small
\caption{The Physical Properties of massive PEGs in our sample \label{tb:ppp}}
\begin{tabular}{ccccccccc}
\hline\hline
{\footnotesize GOODS ID} & {\footnotesize RA} & {\footnotesize DEC} & redshift\footnote[1]{The number in brackets indicate the quality of redshifts: \textit{p} stands for photometric redshift, \textit{s} for spectroscopic redshift} & ${\rm M_{star}}$ & ${\rm SSFR}$ & {\footnotesize E(B-V)} & {\footnotesize Z} & ${\rm R_{eff}}$ \\
          & (J2000) & (J2000) & & ${\rm Log(M_\odot)}$ & ${\rm Log(yr^{-1})}$ & & & kpc \\
\hline
19389 & 53.1357303 & -27.7849320 & 1.345(p) & $10.18\pm0.11$ & $-11.98\pm1.19$ & $0.10\pm0.04$ & ${\rm 1.0Z_\odot}$ & 1.02 \\
22704 & 53.1537988 & -27.7745867 & 1.384(s)  & $10.70\pm0.01$ & $-14.55\pm1.00$ & $0.15\pm0.07$ & ${\rm 0.2Z_\odot}$ & 0.50 \\
23495 & 53.1584550 & -27.7739817 & 2.422(p) & $11.07\pm0.05$ & $-11.98\pm1.54$ & $0.25\pm0.06$ & ${\rm 0.2Z_\odot}$ & $<$0.38 \\
23555 & 53.1588102 & -27.7971545 & 1.921(s)  & $10.82\pm0.04$ & $-11.98\pm0.07$ & $0.00\pm0.01$ & ${\rm 1.0Z_\odot}$ & 0.44 \\
24279 & 53.1630047 & -27.7976545 & 1.980(s)  & $10.63\pm0.07$ & $-12.39\pm0.34$ & $0.00\pm0.01$ & ${\rm 0.2Z_\odot}$ & 0.37 \\
24626 & 53.1651596 & -27.7858696 & 1.317(s)  & $11.10\pm0.04$ & $-11.15\pm0.05$ & $0.10\pm0.03$ & ${\rm 0.2Z_\odot}$ & 3.69 \\
\hline
\end{tabular}
\end{minipage}
\end{table}

\section{Annular Photometry of Massive Passively Evolving Galaxies}
\label{aperphot}

\begin{figure*}[htbp]
\center{\includegraphics[scale=0.9, angle=0]{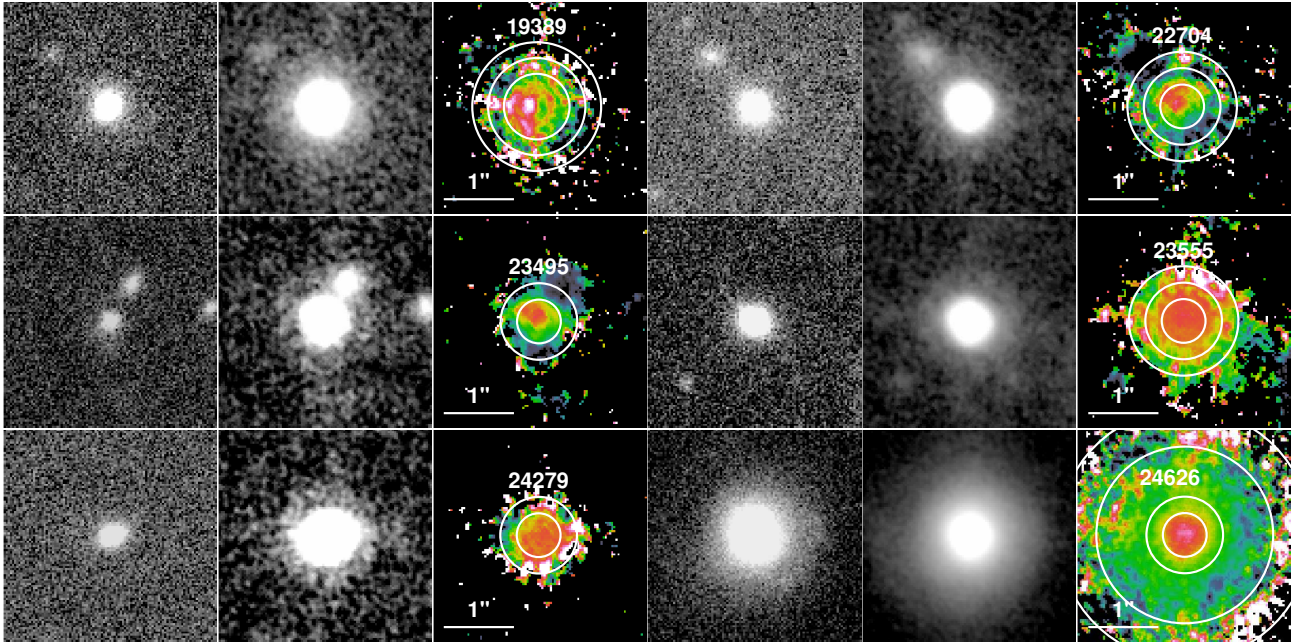}}
\caption[]{The montage of six massive passively evolving galaxies in our
  sample. Each row shows two galaxies. For each galaxy, panels from left to
  right show the HUDF HST/ACS z-band, WFC3/IR F160W, and z-H color images. The
  GOODS v2.0 ID of each galaxy is labeled in images.  The z-band and H-band
  images have different resolution (PSF FWHM of 0.12\arcsec\ and 0.18\arcsec,
  respectively), but the z-H color images are generated after matching the
  z-band PSF to that of H-band (see \S\ref{sub:psfmatch} for details). The
  white concentric circles outline the annuli used to measure the multi-band
  annular photometry. For each galaxy, a white line shows the scale of
  1\arcsec.
\label{fig:mtg}}
\vspace{-0.2cm}
\end{figure*}

We measure azimuthally--averaged color gradients for the six galaxies by
carrying out aperture--matched, multi--band annular photometry. A problem we
face when implementing this procedure was how to properly define the set of
concentric apertures for each galaxy in such a way that it optimally samples
the color gradient. After some experimentation with automated procedures to
determine the annuli based on the effective radius (typically in the H--band),
however, we resort to set them manually based on a visual inspection of the
(z-H) color images. We test the robustness of our result against the choice
of the apertures by perturbing them around the visually determined positions
and also by choosing equally--spaced annuli simply based on the (visually
established) extent of the color gradient. While variations at the level of
10--15\% were observed, in no case these would change our results and
conclusions. The chosen annular apertures for each galaxy are shown as white
circles in Figure \ref{fig:mtg}. Obviously, while with a sample of six
galaxies it is relatively simple to manually set the concentric apertures,
dealing with large samples will require an automated procedure to be
developed. We plan to come back to this problem in a future paper.

Before carrying out the multi--band photometry an additional step was
necessary, namely matching the angular resolution of the images in all the
filters to that of the H band PSF to eliminate any artificial color gradients
introduced by differences in image quality. This varies from
${\rm FWHM}\sim$0.12\arcsec\ in the BViz bands to ${\rm FWHM}\sim$0.18\arcsec\ in the YJH
bands. We will discuss this procedure in the next section.

\subsection{PSF Matched Images}
\label{sub:psfmatch}

We measure the PSF in each band from seven well exposed and
non--saturated stars whose SExtractor stellarity index in the i-band is larger
than 0.98. These stars are used as input to the IRAF DAOPHOT package to
generate an average PSF image in each band. DAOPHOT fits an analytical profile
to the central region within $\sim1\times$ FWHM and adds the averaged
outskirts of the stars to the best-fit profile. Once we build the PSF for
each of the BVizYJ bands, we use the package IRAF PSFMATCH to calculate
a smoothing kernel and convolve each image to match its resolution to that of
the H-band.

\begin{figure*}[htbp]
\center{\includegraphics[scale=0.6, angle=0]{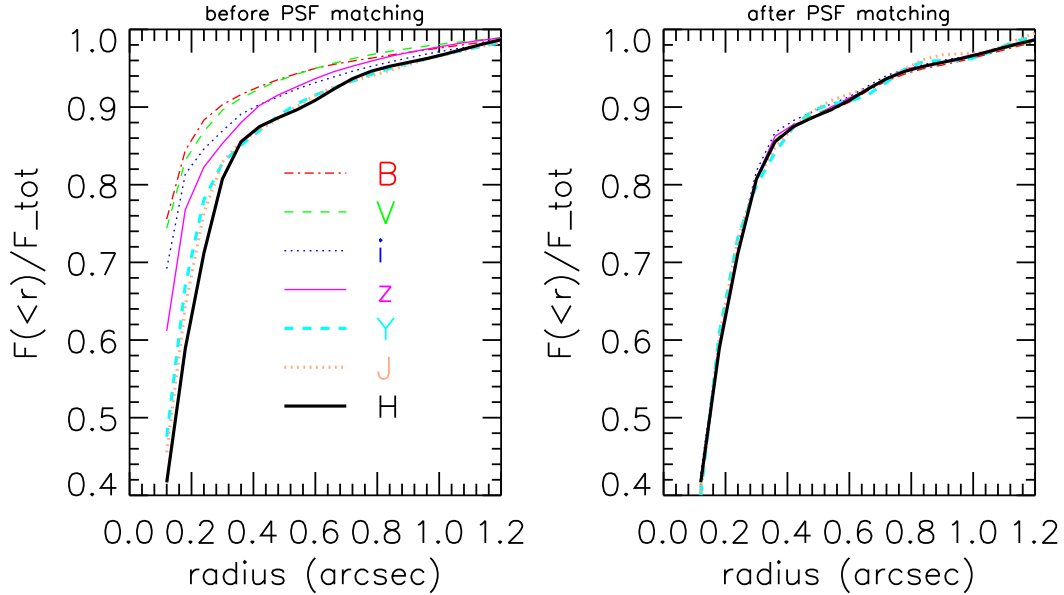}}
\caption[]{The fractional encircled energy of PSFs of 7 bands, before 
({\it left}) and after ({\it right}) our PSF matching.
\label{fig:curveofgrowth}}
\vspace{-0.2cm}
\end{figure*}

We test the effectiveness of PSF matching by comparing the fractional
encircled energy of each PSF before and after the procedure. Figure
\ref{fig:curveofgrowth} shows that after matching, the PSF in all bands have 
identical profile, especially within the central region (roughly
$<$0.4\arcsec\ ), where the gradient is steepest. There are some very small 
fluctuations in the wing (0.4\arcsec\ to 1.0\arcsec\ ) of the Y and J band PSFs,
due to differences in the airy rings of the original PSF. These, however, are 
smaller than 2\% and thus we neglect them, since they will not cause
any detectable bias in our analysis.

\begin{figure*}[htbp]
\center{\includegraphics[scale=0.85, angle=0]{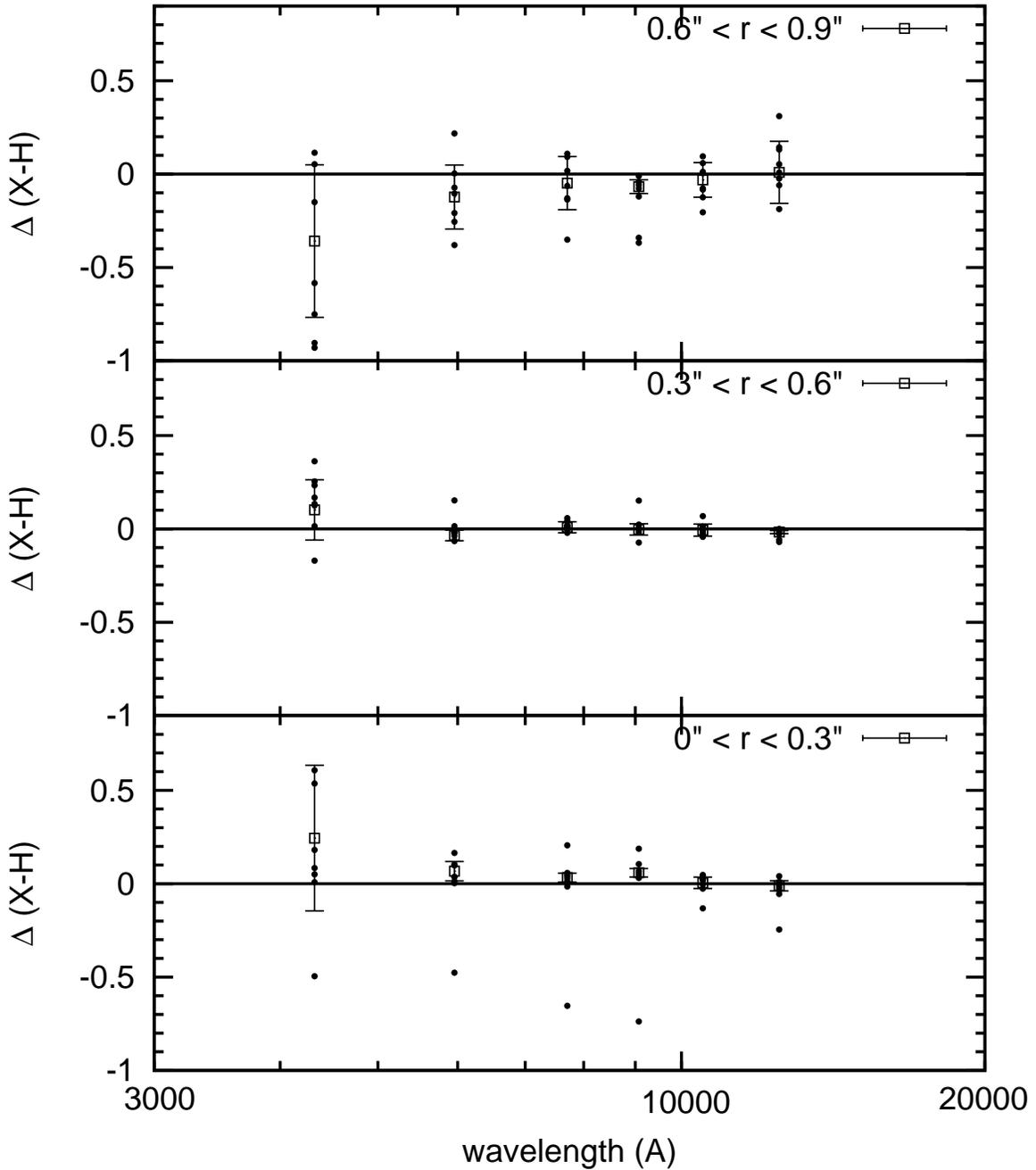}}
\caption[]{The effect of PSF--matching in the measure of color gradients. The
  figure shows the difference ${\rm \Delta (X-H) = output - input}$, where X 
is one of BVizYJ, between the input color gradient of a model galaxy and
the output one, measured from the real images after convolving the model
with the PSF of each image, inserting the result into the image,
applying the PSF--matching procedure and measuring the ``observed'' color
gradient. To simulate the effects of a position--dependent PSF, we do not
use the average PSF of each band, but rather each of the seven stars 
(after appropriate normalization) that we use to create the average
PSF. Thus, each point represent the color gradient of the same model galaxy
observed at different position in the HUDF FOV, while the squares and error
bars show the mean and standard deviation of the points in each band.
\label{fig:simcolor}}
\vspace{-0.2cm}
\end{figure*}

In addition to testing the homogeneity of the matched PSF, we also
verify the effectiveness of the PSF matching procedure in measuring
realistic color gradients by means of simulations that, at the same time, also
give us information on the effects of the PSF variations across the field. We
generate a model galaxy with given colors and then inserted it
into the images in seven different positions In practice, we use each of
the seven stars that went into building the average PSF as position--dependent
PSF themselves, after appropriate normalization. For the model galaxy we 
use a \sersic spheroid with index $n=2$ and effective radius ${\rm R_{eff}}=0.5$ kpc
in the H--band, and assigned (X-H) color, where X is one of BVizYJ. We
convolve the model image with the seven PSFs in each band and inserted the
result in the corresponding image in proximity to the star. Then, we
apply the PSF--matching procedures to the images and measured the color
gradient of the galaxy at its seven different potions as if these were real
measures.  Figure \ref{fig:simcolor} shows the difference ${\rm
  \Delta(X-H)=(X-H)_{out}-(X-H)_{in}}$ between the ``observed'' color gradient
and the input one in each band at each of the seven positions. As the figure
shows, there is no evidence of significant systematic bias introduced by the
PSF--matching procedure, with all the deviations consistent with having random
nature. The case of the B band is the one with the largest deviations, but
while the scatter of ${\rm \Delta(B-H)}$ is comparatively large, the mean
difference between the output and the input colors at radius less than
0.6\arcsec\ does not significantly deviate from zero. In the annulus between
0.6\arcsec\ and 0.9\arcsec, the simulations suggest that we underestimate the
(B-H) color by $\sim0.4$ magnitude, although the B--band flux of PEGs in our
targeted redshift range is so faint that the background fluctuation, rather
than the mismatching of PSFs, likely dominates the uncertainty of the color
measurement. In practice, however, this has no direct consequence in our
analysis, since we do not use B--band derived color gradients. In conclusion,
our test shows that the PSF--matching procedures is effective in recovering
the color gradient and introduces no significant bias to our measurements.

\subsection{The Reliability of the Annular Photometry: the Probability 
Distribution of Photometric Redshift}
\label{sub:photozdist}

We also conduct a further test of the robustness of results derived
from the multi--band annular--aperture photometry by comparing the photometric
redshift derived from each annulus to that measured from the integrated
photometry.  In principle, the redshift of an annulus should be the same as
that of the whole galaxy. If large deviations are encountered this flags
potential bias in results derived from the annular photometry, especially for
the most outer annulus, where the S/N in the bluer bands is significantly
lower than the redder ones and, as we have seen for the B band, other
systematics can affect the measures.

\begin{figure*}[htbp]
\center{\includegraphics[scale=0.6, angle=0]{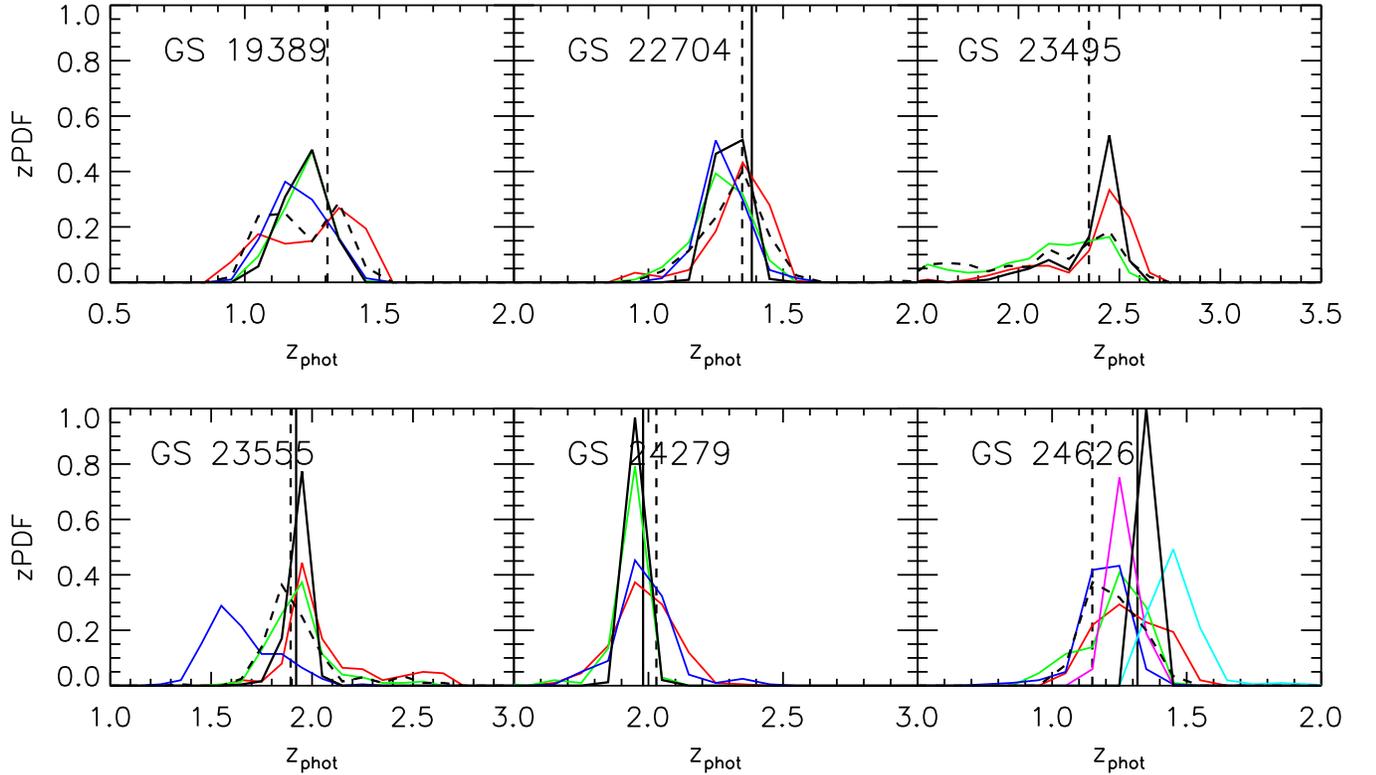}}
\caption[]{The probability distribution function of photometric redshift
  measured (from the {\it HST} BVizJH photometry) in concentric annuli around
  our sample galaxies compared to that of the galaxies as a whole, as well as
  to their spectroscopic redshift, when available. The concentric annuli, from
  the center to the outskirts of each galaxy, are plotted with the red, green,
  blue, violet, cyan, light brown and gray curves. The combined probability,
  i.e. the product of that of each annulus, is also plotted with a black solid
  curve. The black dashed curve shows the probability for each galaxy as a
  whole. The solid vertical line shows the spectroscopic redshift (when
  available), while the dashed vertical line shows the photometric
  redshift of the galaxy measured using the 12-band GUTFIT (integrated)
  photometry.
\label{fig:zpdf}}
\vspace{-0.2cm}
\end{figure*}

Figure \ref{fig:zpdf} shows the probability distribution function of the
photometric redshift measured with the {\it HST} BVizJH photometry for the
annuli and for the whole galaxy for each of our sources. The figure also plots
the photometric redshift of the galaxies derived
from the integrated GUTFIT photometry, as well as the spectroscopic redshift
if available. Generally, there is good agreement between the photometric
redshift of the annuli and that of the whole galaxy, with the differences
between the peaks of the distribution function of the annuli and the whole
galaxy photometric or spectroscopic redshift being typically $\Delta
z/(1+z)<0.05$. Exceptions are two of the annuli of galaxy 24626, which deviate
from the spectroscopic redshift by $\Delta z/(1+z)\sim0.08$, and the outermost
annulus of galaxy 23555, which differs from both the spectroscopic and
photometric redshifts (which agree very well with each other) by $\Delta
z/(1+z)\sim0.12$.

Overall, the agreement between the annuli's photometric redshift and
the spectroscopic or photometric redshift of the whole galaxies is typical of
this types of measures, with no indication that fitting of the observed
SED of the annuli to stellar population synthesis models to derive the
properties of the stellar populations might be affected by systematics or
other problems.

Finally, we wish to point out that the availability of resolved multi--band
photometry of sub--structures with more homogeneous color distribution than
the whole galaxy provides a powerful means to improve the photometric redshift
measurements, as well as to investigate the reason behind catastrophic
failures. Although the redshift probability distribution of the individual
sub--structure does, in general, deviate from the true redshift due to random
errors, the combined probability distribution, i.e. their product, is
generally closer to the true redshift and more sharply distributed than that
of the whole galaxy, because the simpler case of the homogeneous colors of the
sub--structures is better described by the stellar population synthesis
models than the more complex case of the generally much larger color
dispersion inside the whole galaxy. This is illustrated in Figure
\ref{fig:zpdf}, which shows that the peak of the combined redshift probability
distribution of each galaxy is closer than the individual ones to the
spectroscopic or the (12--band) photometric redshift, with typical deviations
$\Delta z/(1+z)<0.03$. For example, although the annuli redshift probability
distribution of galaxy 24626 show relatively large deviations from the
spectroscopic redshift, the combined distribution deviates only by $\Delta z/(1+z)
\sim 0.02$, a more accurate estimate than that of the 12--band photometric
redshift, which has $\Delta z/(1+z)\sim 0.06$. We plan to return on this
technique using data from the {\it HST} CANDELS (Cosmic Assembly Near 
Infra--red Deep Extragalactic Legacy Survey) program (co-PIs: Sandra Faber 
and Henry Ferguson), which in portions of the survey area will include 
photometry in two additional filters, F814W and
F998W, in addition to those discussed here. This will provide even more
accurate estimates of photometric redshift and stellar population parameters.

\section{Color Gradients in Massive Passively Evolving Galaxies}
\label{cg}

\begin{figure*}[htbp]
\center{\includegraphics[scale=0.5, angle=0]{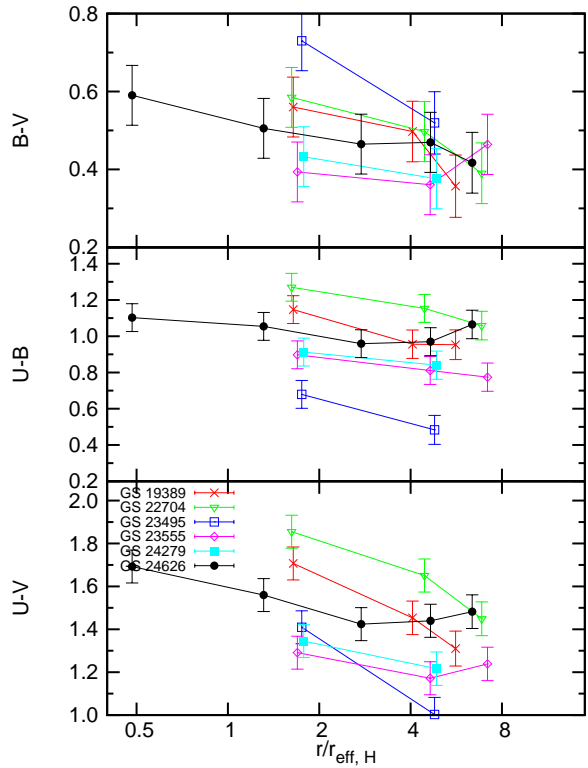}}
\caption[]{The rest-frame B-V ({\it top}), U-B ({\it middle}) and U-V ({\it
    bottom}) color gradients of the six sample galaxies. Each galaxy gradient 
    is color and symbol--coded as labeled in the bottom panel. Also shown are the IDs of 
    the galaxies in the publicly released GOODS v2.0 source catalog. 
\label{fig:cg}}
\vspace{-0.2cm}
\end{figure*}

To investigate possible dependence of the color gradients of the $z\sim 2$
PEGs with other integrated physical properties of the galaxies
and also to compare them to those of local early--type galaxies, we interpolate
the observed photometry in the annuli to the rest--frame U, B and V band and
then obtain the corresponding (U-B), (U-V) and (B-V) colors (e.g. see
\citet{dahlen05}). Figure \ref{fig:cg} shows the color gradients of the
six massive PEGs, where the rest--frame colors of the annuli are
plotted against the annulus radius expressed in unit of the H--band
half--light radius (${\rm R_{eff,H}}$). For five galaxies the available
angular resolution and sensitivity allow us to measure the color
gradients from $\sim1.5\times{\rm R_{eff,H}}$ to $\sim8\times{\rm
  R_{eff,H}}$. For galaxy 24626, due to its much larger size (in the H band we
measure ${\rm R_{eff,H}}\sim3$ kpc), we are able to follow the color
gradients down to a much smaller radius, $\sim0.5\times {\rm R_{eff,H}}$.

To the extent that our sample is representative of early--type galaxies at
$z\sim2$, it appears that these systems have negative color gradients in all
the three colors that we consider, in the sense that stellar population
in these galaxies becomes bluer with increasing separation from the
center. This property can already be inferred from a visual inspection of the
(z-H) color images shown in Figure \ref{fig:mtg}, where all galaxies exhibit
red cores and blue outskirts.

The colors of two of the galaxies appear invert the blueing trend at large
radii, i.e. their color gradient shows an upturn to the red at ${\rm R/R_{eff,H}\approx 3}$--4. Galaxy 23555 exhibits the red upturn in both the (B-V) and (U-V)
color gradients. The photometric redshift probability distribution of the
outermost annulus of this galaxy (see \S\ref{sub:photozdist}) shows a
relatively large deviation from its spectroscopic redshift, suggesting that
the photometry of this area of the galaxy is subject to some systematics. A
visual inspection of the the H--band image reveals that this galaxy resides
in a relatively dense environment, with a luminous, large companion and a
bright star located nearby. Low--surface brightness H--band light from these
sources is very likely contaminating the outermost annulus of the galaxy.
Galaxy 24626 has upturns in the (U-B) and (U-V) color gradients. Although
there are no large or bright sources nearby, a few faint ones are located
close to its outermost annulus. These sources are also more extended in the
near--IR bands than at the optical ones, and may significantly contribute red
light to the outskirts of the galaxy.

Our findings of red cores and blue outskirts in massive PEGs 
at $z\sim 2$ are in apparent contradiction of what reported by
\citet{menanteau01a,menanteau04}, who also find that a large fraction
($\gtrsim$30\%) of spheroidal galaxies at $z\sim0.5$ have strong internal
color variations, but in most of their cases the cores appear bluer than the
surrounding areas, suggesting that blue cores are common in $z\sim0.5$
elliptical galaxies. \citet{menanteau01b} even concluded that most ($\sim$60\%)
of their spheroids formed at $z\lesssim2$. 
Regardless of the difference of targeted cosmic epochs between their and
our works, different sample selection criteria could be the main reason of the 
apparent discrepancy. Our galaxies are selected with both early--type 
morphology and very low SSFR determined by SED--fitting, while 
\citet{menanteau01a,menanteau04} only selected galaxies with 
E/S0 morphology in {\it HST I} F814W images, independent of their potential 
star--formation activity. We also note the a recent work by \citet{gargiulo11}
also reported that 50\% of their sample of 20 early--type galaxies 
at z$\sim$1.5 has significant radial color variation, with five with 
red cores and five with blue cores. Their sample was also selected 
through morphology, mainly based on the visual inspection of {\it HST}/ACS
F850LP images and further cleaned by removing sources with \sersic index n$<$2 
or clear irregular residuals resulting from light profile fitting 
\citet{saracco10}. It is likely that the slope of color gradient (negative
or positive) has relation with the star--formation activity of 
galaxies, even they all have early--type morphology. Besides, both 
\citet{menanteau01a,menanteau04} and \citet{gargiulo11} also found a significant
fraction (40\%$\sim$50\%) of their galaxies to have red cores as ours. 
However, our sample only contains six galaxies and cannot allow to carry out 
a good statistical analysis to compare with them. The upcoming CANDELS will 
provide much larger samples to evaluate the fraction of red cores in 
early--type galaxies at z$\sim$2.
% Regardless of the differences in
% selection criteria and targeted cosmic epochs between our works, we note that
% in their study they elected not to match the PSF of their various bands. The
% very shallow dependence of the apparent angular diameter with redshifts (for
% example, one arc second subtends 6.1 physical kpc at $z=0.5$ and 8.4 at $z=2$, a
% mere 27\% difference) is not sufficient to fully compensate for the very
% pronounced dependence of the encircled energy (EE) of the PSF with
% wavelength. For example, Figure 2 shows $\approx 8$\% difference in the EE
% between z and V, roughly V and B at $z=0.5$, at $r=0.2$ arc second, and as much as
% $\approx 23$\% at $r=0.1$ arc second. Thus, we believe that the sharper PSF of the
% bluer bands inevitably increases the portion of blue light in the center of
% galaxies, resulting in an artificially bluer cores.

\begin{figure*}[htbp]
\center{\includegraphics[scale=1,angle=0]{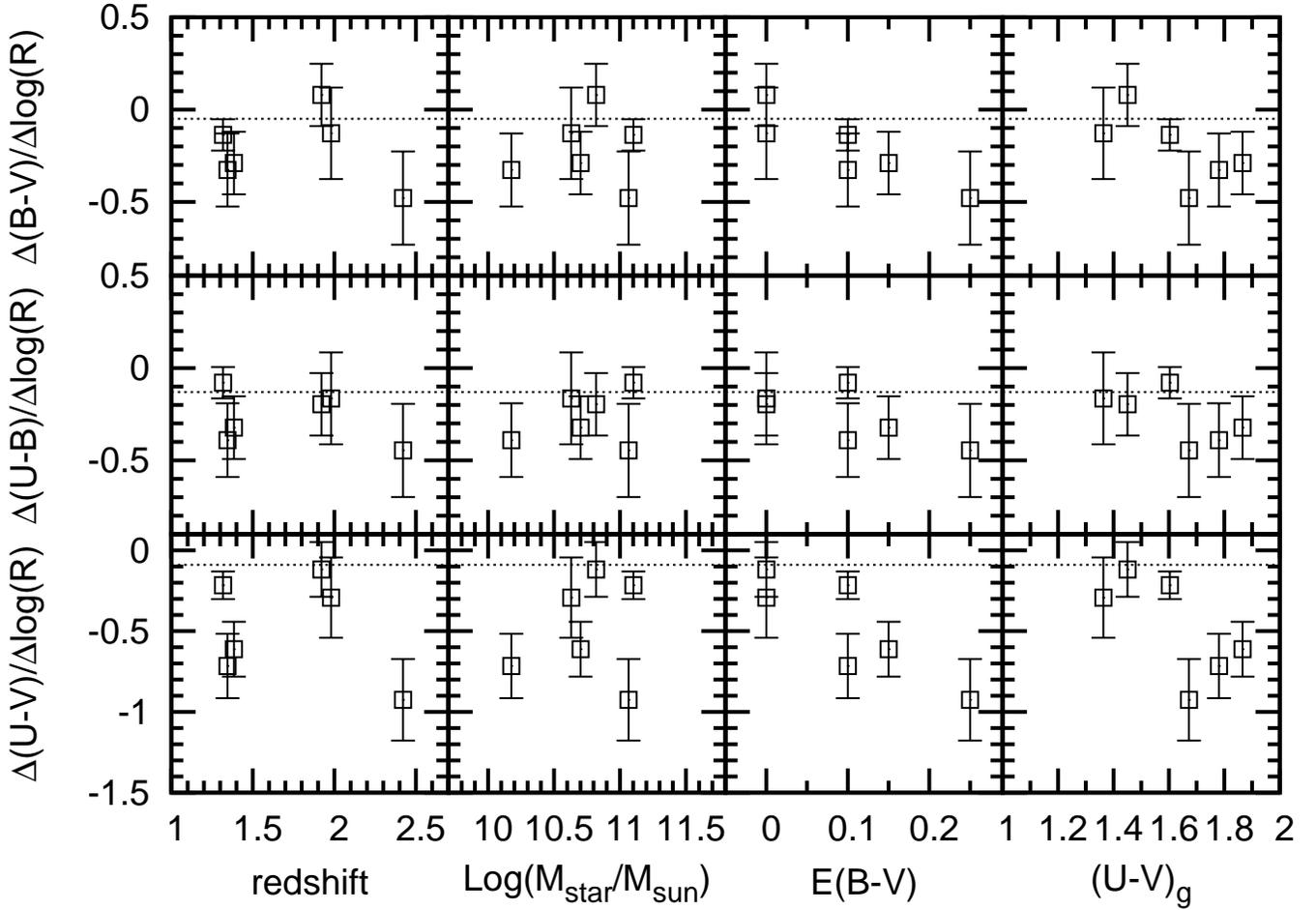}}
\caption[]{The slope of the (B-V) ({\it bottom row}), (U-B) ({\it middle row})
  and (B-V) ({\it top row}) color gradients of the six sample galaxies as a
  function of redshift, stellar mass ${\rm M_{star}}$, E(B-V) as a proxy for
  dust obscuration, and global rest--frame U-V color. The properties of the
  stellar populations of the galaxies are derived from fitting the 12--band
  GUTFIT photometry to spectral population synthesis models, as explained in
  the text. The dotted line in each panel shows the slope of the color
  gradients of local elliptical galaxies measured by \citet{wuhong05}.
\label{fig:cgslope}}
\vspace{-0.2cm}
\end{figure*}

We investigate the dependence of the color gradients on the integrated
properties of the stellar populations of the galaxies. Figure
\ref{fig:cgslope} shows the slope ${\rm \Delta C / \Delta log(R)}$ (C and R
are the color and the radius) of the color gradients as a function of
redshift, stellar mass ${\rm M_{star}}$, color excess E(B-V) as a proxy of
dust obscuration, and the global rest--frame (U-V) color of the galaxies.  The
properties of the stellar populations have been measured from fitting the
12--band GUTFIT photometry of the whole galaxies to spectral population
synthesis models, as described in \S\ref{pegs}. We find that the slopes
have a mild dependence on the the dust extinction E(B-V), in the sense that
galaxies with higher dust obscuration tend to have steeper color gradient
(larger slopes). At face value this seems to suggest that the origin of color
gradients is somehow related to the dust content of the galaxies. We also
find that slopes have a weak dependence on the global rest-frame (U-V) colors of
galaxies, with redder (U-V) colors corresponding to steeper color
gradients. No dependence of the slopes on redshift and ${\rm M_{star}}$ is
could be observed.

We also compare the slopes of the color gradients of the $z\sim2$
galaxies with that of local ellipticals (dashed lines). The local slopes were
measured by \citet{wuhong05}, who studied the color gradients of a sample
of 36 nearby early--type galaxies from the Early Data Release of the Sloan
Digital Sky Survey and from the Two Micron All Sky Survey. The slopes of the
$z\sim2$ galaxies that have little or no dust extinction are similar to those
of the local galaxies, while the $z\sim2$ galaxies with more pronounced
obscuration have steeper color gradients. The color gradients of local
elliptical galaxies are generally interpreted as evidence of metallicity
gradients \citep[e.g.,][]{tamura00,wuhong05, labarbera09}. We will investigate
the origins of the color gradients in the $z\sim2$ galaxies in next two
sections.

\section{Variation of Single Parameter as the Origin of Color Gradients}
\label{ssp}

In view of the analysis of the color gradients with SED fitting to spectral
population synthesis models to understand their physical origin, in this
section we investigate whether it is plausible that the radial variation of
one single parameter can be primarily responsible for them. In other words,
the observed color gradients can, in general, be explained to the radial
variation of age, dust obscuration and metallicity of the stellar populations,
either individually or in combination. Here we constrain the radial gradient
of any one of these parameters needs to be, while keeping the others
constant, for it to be solely responsible for the observed color gradients and
discuss the implications. We assume simple parametrization for the dependence
of the selected parameter with the radius, and, for simplicity, we only use
a single stellar population (SSP) model from CB09 as representative of the SED
of our passively evolving galaxies.

\begin{figure*}[htbp]
\center{\includegraphics[scale=0.8, angle=0]{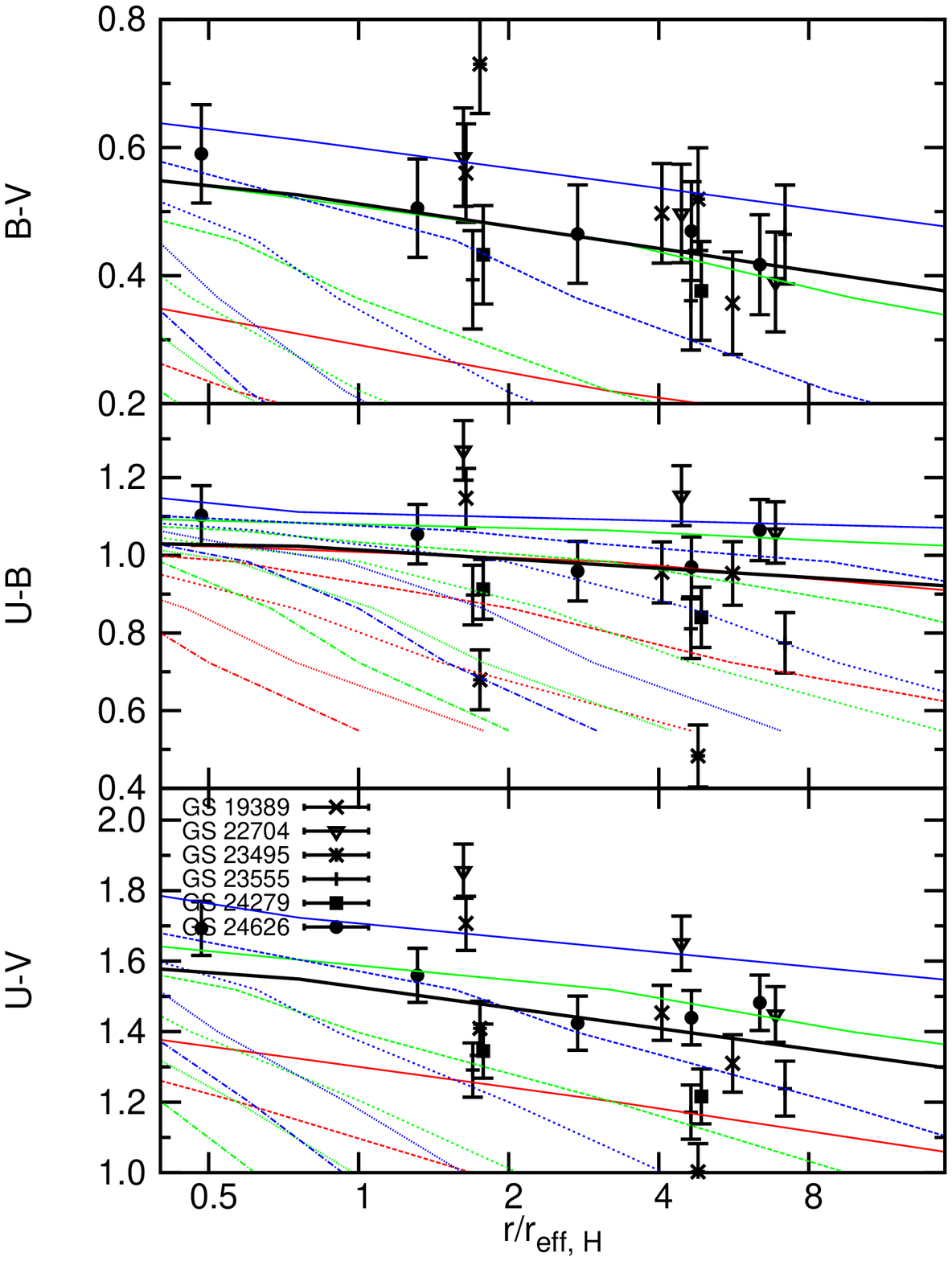}}
\caption[]{The predicted rest--frame (B-V), (U-B) and (U-V) color gradients
  from single stellar population models if the age of the dominant stellar
  population is the only parameter that varies as a function of radius in the
  galaxies (thin lines). The observed colors are plotted as black symbols with
  error bars.  We only plot the predictions for the case of solar metallicity
  and zero dust extinction in this figure. Blue, green and red lines show the
  models in which the age at the center has been set at 3, 2 and 1 Gyr,
  respectively. For each color, different line patterns show different age
  gradients models, i.e., from top to bottom, ${\rm \Delta log(age)/\Delta
    log(R/R_{eff})}$ =-0.2 (solid), -0.4 (long dashed), -0.6 (short dashed),
  -0.8 (dotted) and -1.0 (dashed-dotted). The think black line shows the
  prediction of age gradient that best reproduces the observations. See the
  text for details. 
\label{fig:dage}}
\vspace{-0.2cm}
\end{figure*}

First, we study the possibility that an age gradient is responsible for the
observed color gradients, while keeping metallicity and dust obscuration
constant with radius. We model the age gradient as \\
${\alpha_t = {\rm \Delta log(t) / \Delta log(R/R_{eff})}}$,
where $t_0$ is the age at the center,
and thus a model is fully described by a set of four values of the parameters
$t_0$, $\alpha_t$, Z, E(B-V). Rather than finding best--fit models we let the
parameters vary within a four dimensional (4--D) grid chosen so that the results
model predictions for the color gradients bracketed the observed ones. The 4--D
grid is defined by ${\rm Z = 0.2Z_\odot}$, ${\rm Z_\odot}$ and ${\rm
  2.5Z_\odot}$, E(B-V) = 0.0, 0.05, 0.10 and 0.15, $t_0$ = 1, 2 and 3 Gyr,
${\alpha_t}$ = -0.2, -0.4, -0.6, -0.8 and -1.0. Given a point in the grid,
i.e. the vector ($t_0$, $\alpha_t$, Z, E(B-V)), we compute the color gradients
of (U-B), (U-V) and (B-V) as a function of radius. We also 
calculate the $\chi^2$ as a metric to characterize the goodness of a model in
describing the observations, defined as:
\begin{equation}
  \chi^2 = \frac{1}{N_{obs}} \sum_{i=1}^{N_{obs}} \frac{(C_{obs,i}-C_{p,i})^2}{\sigma_{C,obs,i}^2}, 
\label{eq:chi2}
\end{equation}
where $C_{obs,i}$ and $\sigma_{C,obs,i}$ are the observed color and its
uncertainty at a given radius, $C_{p,i}$ the predicted color at the radius,
and $N_{obs}$ the total number of observed colors at all radii.

Figure \ref{fig:dage} shows the model color gradients for (B-V), (U-B) and
(U-V) compared them with the data. For simplicity, we only show the case of
Z=${\rm Z_\odot}$ and E(B-V)=0 in the plot. Blue, green and red lines
correspond to $t_0$ = 3, 2 and 1 Gyrs, while the different line patterns show
the cases of $\alpha_t$: -0.2 (solid), -0.4 (long dashed), -0.6 (short
dashed), -0.8 (dotted) and -1.0 (dashed-dotted).

The {\it top} panel shows that the (B-V) color gradient is best approximated
by the solid green line, i.e. $t_0=2$ Gyr and $\alpha_t$=-0.2
($\chi^2=1.23$). But the {\it middle} and {\it bottom} panel show that this
set of parameters overestimates the (U-B) ($\chi^2=7.73$) and (U-V)
($\chi^2=7.12$) color gradients. The (U-B) color gradient is actually best
approximated by the green long-dashed line ($t_0=2$ Gyr and $\alpha_t$=-0.4,
$\chi^2=5.42$), while the (U-V) one by the long--dashed blue line ($t_0=3$ Gyr
and $\alpha_t$=-0.4, $\chi^2=5.71$).

Even when we change value of the Z and E(B-V) within the preassigned range we
still cannot find a combination of $\alpha_t$ and $t_0$ that can simultaneously
provide a good description for all the three color gradients. The parameter
set that best reproduces the (U-B) color gradient is ($t_0$, $\alpha_t$, Z,
E(B-V)) = (3.0, -0.6, 0.2${\rm Z_\odot}$, 0.15) with $\chi^2=5.31$, that of
(U-V) by (3.0, -0.4, ${\rm Z_\odot}$, 0.0) with $\chi^2=5.71$, and that of
(B-V) by (1.0, -0.2, 2.5${\rm Z_\odot}$, 0.15) with $\chi^2=1.10$.
We also determine which model minimizes the combined $\chi^2$, namely
$\chi^2_{U-B}+\chi^2_{U-V}+\chi^2_{B-V}$. This model, shown by the black lines
in the figure correspond to ($t_0$, $\alpha_t$, Z, E(B-V)) = (3.0, -0.2,
0.2${\rm Z_\odot}$, 0.0), with $\chi^2_{U-B} = 5.58$, $\chi^2_{U-V} = 5.89$
and $\chi^2_{B-V} = 1.19$.

Since there always is a different combination of the model parameters in our chosen 4--D
grid that brackets different set of observed colors, we conclude that no
combination of $\alpha_t$ and $t_0$ with constant Z and E(B-V), i.e. age
alone, can simultaneously explain the three observed gradients.

We also repeat the same analysis for the case of a metallicity gradient
and obscuration gradient to see if either one of these could be responsible
for the color gradients, finding similar negative conclusions. 

For the case of the metallicity gradient, the parameter sets that best fit the
(U-B), (U-V) and (B-V) color gradients are ($Z_0$, $\alpha_Z=\Delta log(Z) /
\Delta log(R/R_{eff})$, t, E(B-V)) = (2.5${\rm Z_\odot}$, -0.4, 1.0, 0.0) with
$\chi^2=5.30$, (${\rm Z_\odot}$, -0.6, 1.0, 0.15) with $\chi^2=5.44$ and
(2.5${\rm Z_\odot}$, -0.8, 1.0, 0.15) with $\chi^2=1.08$. The corresponding
parameter set that results in the minimum combine $\chi^2$ is (${\rm Z_\odot}$,
-0.6, 1.0, 0.15) with $\chi_{2,U-B} = 5.39, \chi_{2,U-V} = 5.44, \chi_{2,B-V}
= 1.18$.  For the case of the obscuration gradient, the parameter sets that
best reproduce the (U-B), (U-V) and (B-V) color gradients are ($E(B-V)_0$,
$\alpha_{E(B-V)}=\Delta E(B-V) / \Delta log(R/R_{eff})$, t, Z) = (0.15, -0.08,
1.0, 0.2${\rm Z_\odot}$) with $\chi^2=5.49$, (0.15, -0.04, 1.0, 0.2${\rm
  Z_\odot}$) with $\chi^2=6.70$ and (0.15, -0.08, 1.0, ${\rm Z_\odot}$) with
$\chi^2=1.22$. The one that minimizes the combined $\chi^2$ is (0.15, -0.06,
1.0, 0.2${\rm Z_\odot}$) with $\chi_{2,U-B} = 5.53, \chi_{2,U-V} = 6.99,
\chi_{2,B-V} = 2.12$.

In conclusion, unless the assumption of SSP is grossly inadequate for
describing the rest--frame UV/Optical SED of our passively--evolving massive
galaxies at $z\sim 2$, it seems unlikely that the radial dependence of only
one parameter among age, metallicity or dust obscuration (with the other two
being constant) can be responsible for the observed color gradients. These
must originate from the interplay of the gradients of age, extinction and
metallicity.

% \begin{figure*}[htpb]
%   \begin{center}
%     \begin{tabular}{ccc}
%       \resizebox{60mm}{!}{\includegraphics{./plot/restcgsspdage_aper_pix6_peglx_good_psfmatch3_swrms2_2_psfre_log.eps}} &
%       \resizebox{60mm}{!}{\includegraphics{./plot/restcgsspdz_aper_pix6_peglx_good_psfmatch3_swrms2_2_psfre_log.eps}} &
%       \resizebox{60mm}{!}{\includegraphics{./plot/restcgsspdebmv_aper_pix6_peglx_good_psfmatch3_swrms2_2_psfre_log.eps}} \\
%     \end{tabular}
%     \caption{This is sample figures.}
%     \label{test4}
%   \end{center}\label{haha}
% \end{figure*}

\section{Stellar Population Gradients in Massive Passively Evolving Galaxies}
\label{spg}

\begin{figure*}[htbp]
\center{\includegraphics[scale=0.6, angle=0]{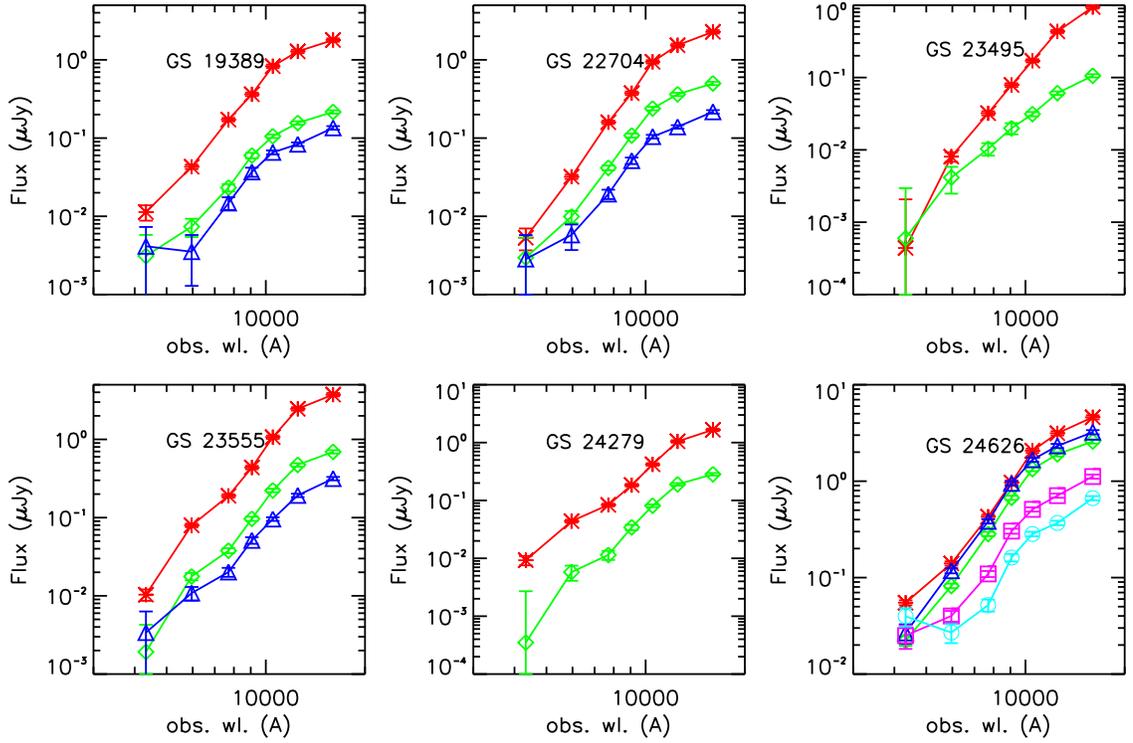}}
\caption[]{The {\it HST} 7--band photometry (ACS BViz and WFC3/IR YJH) of the
  sample galaxies in the annular apertures discussed in Section 4. The curve
  for each annulus is color and symbol--coded as red star, green diamond, 
  blue triangle, violet square, and cyan circle in
  going from the center to the outskirts of each galaxy. 
\label{fig:obssed}}
\vspace{-0.2cm}
\end{figure*}

We investigate the nature of the observed color gradients by fitting the
{\it HST} 7--band photometry (ACS BViz and WFC3/IR YJH) in the annular
apertures defined before (see Figure \ref{fig:obssed}) to the CB09 spectral
population synthesis models to derive the radial dependence of stellar mass,
specific star--formation rate, age and dust obscuration of the stellar
populations in the annuli. We approximate the star formation history 
with an exponentially declining model ($e^{-t/\tau}$), where the age
of the stellar populations is the time $t$ from the beginning of the star
formation to the time of observation. During the fitting, the redshift of
each annulus is kept fixed to the spectroscopic redshift or to the 
photometric redshift of the whole galaxy measured from the GUTFIT 12--band
photometry. 

While this procedure yields robust estimates of the stellar mass, dust
obscuration, age and metallicity suffer from larger uncertainties and
degeneracies \citep[e.g.,][]{papovich01,shapley01,joshualee10,maraston10}.
% (see Papovich et~al. 2001; Shapley et al.~2001; ..., Lee et~al. 2010; Maraston et~al. 2010). 
The degeneracy between age, metallicity
and dust obscuration is partially broken if rest--frame infrared photometry is
available, as shown by several authors
\citep{dejong96d,cardiel03,macarthur04b,wuhong05}.  Unfortunately,
high--angular resolution photometry for our galaxies is limited to rest--frame
UV and optical wavelengths, and thus we cannot effectively separate the role
that each parameter plays in the observed color gradients. To gain some
insight, however, we can make some simplifications and reduce the number of
free parameters. Instead of letting the metallicity free to vary in each
annulus during the fit, we set it according to one of the following three
assumed power--law metallicity gradients: (1) flat, with logarithmic slope
${\rm \Delta log(Z) / \Delta log(R) = 0.0}$; (2) the metallicity gradient of
local early--type galaxies, with ${\rm \Delta log(Z) / \Delta log(R) = -0.25}$
\citep{wuhong05}; (3) the gradient predicted by the monolithic collapse model,
with ${\rm \Delta log(Z) / \Delta log(R) = -0.5}$ \citep{carlberg84}. The
latter model is meant to represent the case where the $z\sim 2$ galaxies have
formed ``in situ'' at some epoch prior that of observation through some
relatively rapid process.

In the local universe, elliptical galaxies have very little dust obscuration
and their stellar populations are essentially coeval in the sense that their
age spread is small compared to the mean age \citep[e.g.,][]{tamura04,wuhong05,labarbera09}. The situation can be very
different at $z\sim 2$. The universe is only $\approx 3.5$ Gyr old at this
time, and thus the approximation of coevality is almost certainly no longer
valid, since this time is comparable to that required to make a galaxy develop
an early--type SED following the cessation of star formation. Furthermore, we
do not understand the mechanisms of dust destruction well enough to make
robust predictions on the dust content of early--type galaxies at $z\sim
2$. Dust is expected to disappear on a time--scale of $\sim 10^8$ years after the
end of star formation, but this is not observed \citep[e.g.,][and reference therein]{draine09}. Thus, 
% while we have included the case of no dust in our analysis, 
we study the more general case where both dust
obscuration and age are left as free parameters. We will discuss the case of 
no dust in our analysis later.
%\includegraphics[totalheight=4cm,angle=0,origin=c,scale=1.5]{./plot/rdist_age_mean_chab_aper5_cen_rr_3a.ps}%
%\includegraphics[totalheight=4cm,angle=0,origin=c,scale=1.5]{./plot/rdist_age_mean_chab_aper5_cen_rr_32.ps}%
%\includegraphics[totalheight=4cm,angle=0,origin=c,scale=1.5]{./plot/rdist_age_mean_chab_aper5_cen_rr_3fb.ps}%
%
%\includegraphics[totalheight=4cm,angle=0,origin=c,scale=1.5]{./plot/rdist_age_mean_chab_aper5_cen_rr_fba.ps}%
%\includeg to represent graphics[totalheight=4cm,angle=0,origin=c,scale=1.5]{./plot/rdist_age_mean_chab_aper5_cen_rr_fb2.ps}%
%\includegraphics[totalheight=4cm,angle=0,origin=c,scale=1.5]{./plot/rdist_age_mean_chab_aper5_cen_rr_fbfb.ps}%

\begin{figure*}
\center{\includegraphics[scale=0.6, angle=0]{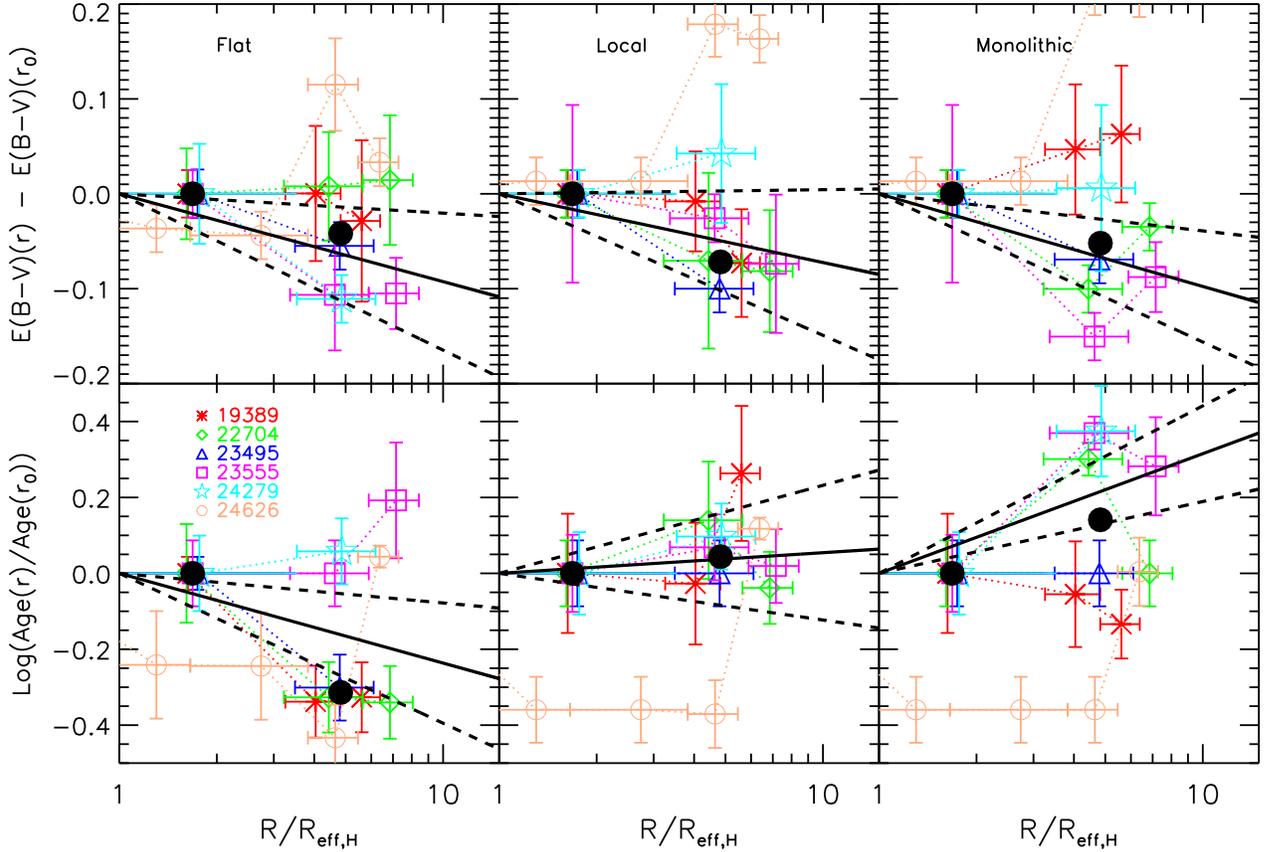}}
\caption{The dust ({\it top} row) and age ({\it bottom} row) gradients of
  massive PEGs under three assumptions of metallicity gradients: (1) the flat
  gradient ({\it left} column), (2) the local gradient ({\it middle} column),
  and (3) the monolithic gradient ({\it right} column).  Galaxies are plotted
  in different colors and symbols, as their IDs show. Horizontal error bars show the size
  of each annulus, while vertical error bars show the 1-$\sigma$ uncertainty
  of each parameter, which is measured through fits on 200 times Monte-Carlo
  sampled SEDs. In each panel, the black solid and black dashed lines show the
  best-fit value and 1-$\sigma$ uncertainty of the slope of gradient of the
  parameter. The two black points in each panel show the median values for the
  bins of ${\rm R/R_{eff}<3.0}$ and ${\rm 3.0<R/R_{eff}<10.0}$.}
\label{fig:covar}
\end{figure*}

Figure \ref{fig:covar} plots, for each galaxy, the gradients of E(B-V) and age
from the fits expressed as the ratio between the value at center and that in
each annulus for each of the three assumptions on the metallicity
gradient. The error bar for each annulus is the standard deviation of the
best--fit values from 200 realizations from Monte Carlo bootstrap
simulations. The figure also shows the average gradient of each parameter and
its best--fit slope $\alpha = {\rm \Delta P / \Delta log(R)}$, where ${\rm P}$
is either E(B-V) or log(age), ${\rm R}$ is the radius, and the average
includes all sample galaxies but 24626. The best--fit slope and its 1$\sigma$
uncertainty are plotted in the figure as black solid and dashed lines,
respectively.

Galaxy 24626 is excluded from the average, because, as Figure \ref{fig:covar}
shows, its gradients of dust obscuration and age are very different from those
of the other galaxies. Its half--light radius and \sersic index, ${\rm R_{eff}\sim 3.7}$ kpc and \sersic index $n=7.4$ \citep[see][]{cassata10}, 
the largest size and most concentrated light profile in the sample, 
as well as its stellar mass, ${\rm Log(M/M_{\odot})}=11.1$,
are typical of the bright elliptical galaxies in the
local universe often observed in groups with estimated total (dark matter)
mass $M\sim {\rm 10^{13} M_\odot}$ \citep{ycguo09cen}. Thus, it is likely
that the star--formation and/or stellar--mass assembly history of this galaxy
considerably differ from those of the other five samples galaxies, a fact that
might reflect in the radial gradients of its stellar population
properties. We also note that although the zYJH band images show a regular
spheroidal morphology out to $\approx 2.5$\arcsec, the B--band image reveals  
that the galaxy has a close companion at about 1\arcsec\, corresponding to 
8.4 kpc or $\approx 2.3\times {\rm R_{eff}}$, from its center. 

Figure \ref{fig:covar} also shows that, regardless of the assumptions on the
metallicity gradient, the implied average E(B-V)s always has a mild gradient
in the sense that the centers (${\rm R/R_{eff}<3.0}$) of the galaxies have
slightly higher dust extinction (${\rm \Delta E(B-V) \sim 0.05}$) than the
outer regions (${\rm 3.0<R/R_{eff}<10.0}$). Both the slope and the amplitude
of the dust gradient do not depend on the assumed metallicity gradient,
implying that a mild negative gradient of dust obscuration is very likely a
real feature of massive PEGs at $z\sim2$, contributing at least in part, to
the observed color gradients. This is consistent with the finding, discussed
in \S\ref{cg}, that the slope of color gradient of the individual galaxies
correlate with the global E(B-V) value, i.e. the one from the best--fit of
each galaxies' GUTFIT 12--band photometry to spectral population synthesis
models.

Due to the age--metallicity degeneracy, however, the contribution of an age
gradient to the observed color gradient is much harder to determine, since it
strongly depends on the assumed gradient of metallicity. As Figure
\ref{fig:covar} shows, a flat metallicity gradient results in the outer
regions of the galaxies being $\sim$60\% younger than the center, while if the
local metallicity gradient were assumed then the galaxies would have a flat
age gradient. Finally, in the case of the metallicity gradient predicted be
the monolithic collapse model, the stellar populations in the outer regions
would by $\sim$2 times older than those in the center. 

Color gradients and internal color dispersion of intermediate--redshift 
early--type galaxies
have been extensively studied in the past \citep[e.g.][]{abraham99,
  menanteau01a, menanteau04}. \citet{abraham99} studied eleven 11 early-type
galaxies at $z\sim0.5$ in the Hubble Deep Field (HDF), finding that most
(7/11) have internal color dispersion consistent with being old and coeval,
and implying a small age gradient in the early--type galaxies at higher
redshift. 
% They argued that the internal color dispersion of these galaxies is
% unlikely to be explained by metallicity gradients, and that the role of dust
%in the color dispersion is minimal. 
Similar properties remain valid at $z\sim 0$ as well \citep{wuhong05}. 
While this is qualitatively consistent with the
null age gradient of our galaxies under the assumption of local metallicity
gradient, in practice a quantitative comparison requires an accuracy in
measuring the age that we do not have. At $z\sim 2$ the age of the universe is
about 3.2 Gyr, while it is 8.4 Gyr at $z\sim 0.5$, and our finding of 
$\sim 1$ Gyr age gradient with flat metallicity gradient means that the 
fractional age differential is $\approx 30$\%. This, however, becomes $\approx 12$\% (or $\approx 7$\% at $z\sim 0$)
just because the universe has become older. In the next section we will
discuss the implications of the assumptions on the metallicity gradients for
the evolution of the galaxies from $z\sim 2$ to the present.

\section{Discussion}
\label{implication}

\subsection{Dust Gradient}
\label{sub:dustgrad}

\subsubsection{Necessity and Robustness}
\label{subsub:dgn}

The mild gradient of dust obscuration, together with its apparent robustness 
against assumptions on the metallicity gradient, that seems to characterize 
massive early--type galaxies at $z\sim 2$ is in general agreement with the 
fact that dust obscuration in early--type galaxies in the local universe is 
not a dominant effect in determining their rest--frame UV/Optical color 
and color
gradient. Thus, it appears that the lack of a significant presence of dust, or
at least of its effects in the UV/Optical rest--frame SED, is a common feature
of passively--evolving galaxies, regardless of the cosmic epoch when they are
observed. Evidently, whatever physical mechanism is responsible for the
destruction of dust in the aftermath of the cessation of star formation in
these systems, must act on a significantly shorter time scale than that
required to make the galaxy's SED become typical of a ``red and dead'' system.

Both the inferred dust gradient and the absolute value of dust obscuration are
comparatively small, ${\rm \Delta E(B-V)/\Delta log (R)\sim-0.07}$ and
$<E(B-V)>\sim 0.1$, and, as we have seen, robust against the assumptions on the
metallicity gradient. An important question is whether or not the opposite is
also true, namely that dust can be neglected when studying the effects of the
age and metallicity gradients in the observed color gradients and their
implications on the evolution of the galaxies, both prior and subsequent to
the epoch of the observations. To answer this question we re--run the
fitting procedure in each annulus under the two assumptions: (1)
zero dust extinction; (2) dust extinction in each annulus fixed to the
integrated value for the whole galaxy from the best--fit of the 12--band
GUTFIT photometry to the models, i.e. no E(B-V) gradient. For simplicity, we
only consider the case of the local metallicity gradient.

\begin{figure*}[htbp]
\center{\includegraphics[scale=0.5, angle=0]{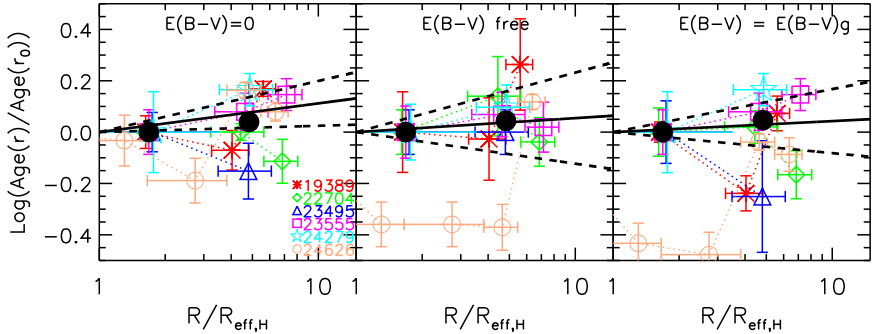}}
\caption{The age gradient of the sample galaxies under three different
  assumptions of dust distribution: (1) no dust; (2) dust as a free parameter
  in each annulus; (3) dust fixed to the global value from the best--fit of
  the 12--band GUTFIT photometry to the models. Individual galaxies are
  color and symbol--coded, as labeled . The horizontal error bars represent the width of
  each annulus, while the vertical ones are the 1-$\sigma$ uncertainty of each
  parameter measured from 200 bootstrap Monte--Carlo realizations of the
  observed SEDs. In each panel the black solid and black dashed curves show
  the best--fit average gradient and 1-$\sigma$ interval. The two black points
  in each panel are the median in the two bins ${\rm R/R_{eff}<3.0}$ and ${\rm
    3.0<R/R_{eff}<10.0}$. The local metallicity gradient is assumed
  throughout.}
\label{fig:dpn}
\end{figure*}

The results are shown in Figure \ref{fig:dpn}, where the new derived age
gradients ({\it left} and {\it right} panels) are compared with the age
gradient derived by letting dust as a free parameter in the fitting 
({\it middle}). While the individual
points vary, albeit within their $1-\sigma$ error bars, the average age
gradient remains unchanged in all three cases, regardless of the assumption on
dust obscuration. 

When the dust obscuration is fixed to zero or to the global 12--band value, the
fit yields systematically larger reduced $\chi^2$ than in the case of free
dust. This is not just the effect of an extra free parameter in the fit: when
we compare the best--fit SED models with the observed photometry, the free
dust case obviously yields better agreement, especially in the B band, the
most affected by dust obscuration. So, dust does play a role in determining
the colors of the galaxies (the dust gradient slope ${\rm \Delta E(B-V)/\Delta
  log (R) \sim -0.07}$ means that the rest--frame B-V color becomes on average
bluer by 0.07 mag from $R_{eff}$ to ${\rm 10\times R_{eff}}$). The absolute value of
dust obscuration and its spatial gradient are small enough, however, that
uncertainties on both these quantities can be neglected when setting
constraints to the age and metallicity gradient from the observed color
gradient, which are important to infer the evolutionary history of the galaxies
as we are going to discuss in the next section.

\begin{figure*}
\center{\includegraphics[scale=0.45, angle=0]{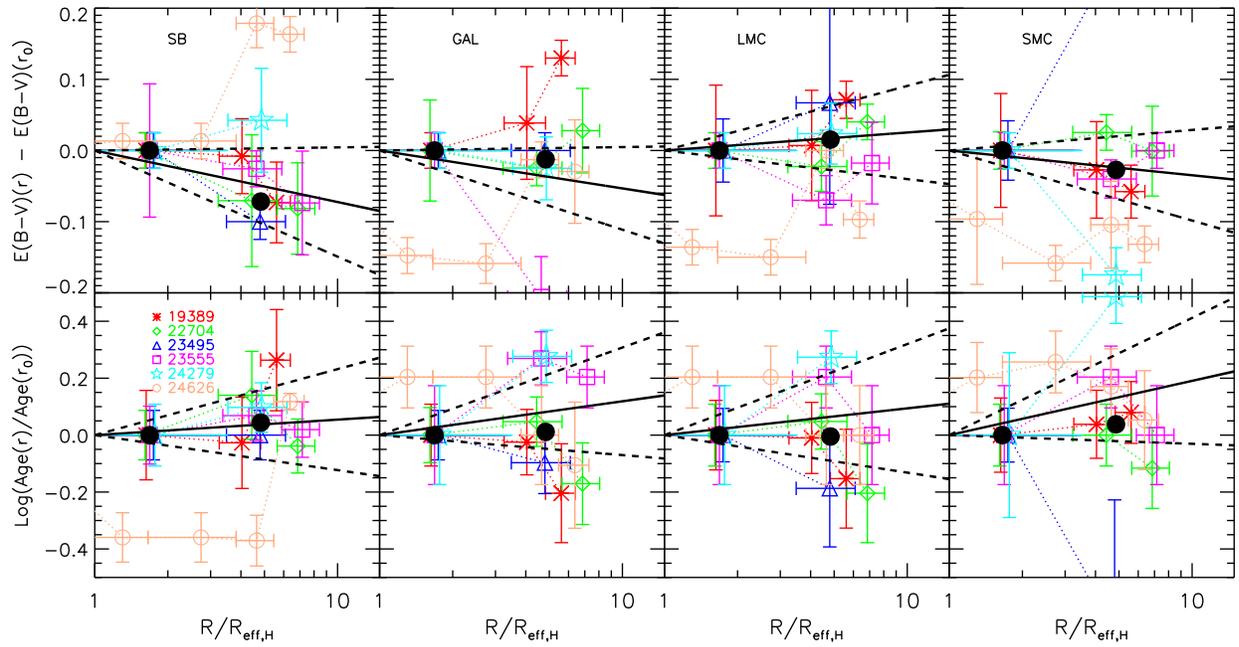}}
\caption{Similar to Figure \ref{fig:covar}, but showing the results of
  different extinction laws.  Panels from left to right show the result of
  Calzetti Law, Galactic Law, LMC Law and SMC Law. The metallicity gradient is
  assumed to be the local one.}
\label{fig:extcn}
\end{figure*}

Throughout this study we have used the starburst obscuration law
\citep{calzetti94,calzetti00} to model the effects of dust. While this is
appropriate in the case of starburst galaxies, it is now known if it remains a
good description in the case of the low SSFR, massive galaxies observed at
$z\sim 2$, such as our sample. Regardless, however, this choice appears to
actually be a conservative one for the implied effects of dust in the color
gradients, as we directly verify by repeating our analysis using the
Galactic \citep{cardelli89}, LMC \citep{fitzpatrick86} and SMC
\citep{prevot84} obscuration laws. For simplicity, we only consider the
case of local metallicity gradient. As Figure \ref{fig:extcn} shows, the slope
of the dust obscuration gradient is reduced to ${\rm \Delta E(B-V)/\Delta log
  (R) \lesssim -0.04}$ when the Galactic and SMC laws are used, and it is
close to zero in the LMC case. 

The color gradient of local early--type galaxies is explained in terms of a
metallicity gradient \citep[e.g.,][]{wuhong05}. As discussed by
\citet{wise96}, however, a dust gradient can also reproduce the observed
broadband color gradients in many ellipticals. Using {\it HST} images,
\citet{vandokkum95} found that 48\% of 64 early-type galaxies show highly
concentrated dust absorption at the centers of the galaxies. The sizes of the
dust absorption regions are generally smaller than 1 kpc. \citet{rest01} and
\citet{tran01} found dust features in 29 out of 67 galaxies (43\%), including
12 with small nuclear dusty disks, while \citet{lauer05etg5} found central
dust in about half of the 77 galaxies that they observed with the {\it
  HST}/WFPC2. 
%\citet{pahre04} found that 3 of their 6 early-type galaxies that
%are observed by SPITZER exhibit dust features within $\sim$2 kpc from the
%centers.

Both an external and internal origin of the dust in the local galaxies have
been proposed. The strongest evidence for the external origin, in which
galaxies obtain their dust from mergers or accretions, is that the
distribution and motions of ionized gas and dust in some locale early-type
galaxies seem unrelated to the motions of stars
\citep[e.g.,][]{goudfrooij95,vandokkum95,caon00}.  However,
\citet{mathews03araa} and \citet{temi07} argued that the dynamical infall time
from the edge of a local early-type galaxy (several $\sim 10^8$ yrs) is
compatible to the time scale of dust destruction due to the sputtering by hot
X-ray gas ($\sim 10^8$ yrs). Therefore, in the external origin, cold gas and
dust should be regularly supplied to galaxies in a time scale of $\sim 10^8$
yrs. But observations of local early-type galaxies do not find such
evidence. In fact, mergers between early-type galaxies and dusty galaxies are
rarely observed in local universe. The lack of effective dust resupplies
strongly points to the internal origin, in which dust is produced inside the
galaxies by either the mass loss from evolving red giant stars
\citep[e.g.,][]{temi07} or M-star winds, as discussed in \citet{lauer05etg5}.

Understanding the origin of dust in massive PEGs at $z\sim2$ requires
spatially resolved stellar and gas kinematics, which are not
available. If they have undergone merger or accretions to some extent in
their past, dust obtained during these events should settle to the center with
dynamical infall time, a few $\sim 10^8$ yrs. If this dust is responsible for
the observed obscuration gradients, then this would argue against the
existence of hot X--ray emitting gas, which would otherwise sputter the dust
in a shorter time scale ($\sim 10^8$ yrs). If such gas does exist, then a more
plausible mechanism to form dust gradients is the episodic settling model
proposed by \citet{lauer05etg5} to explain the existence and frequency of
nuclear dust in local early-type galaxies. The model predicts that dust
appears several time throughout the galaxy and then is destroyed as it falls
into the center. Therefore, the existence of hot gas in the massive PEGs at
$z \sim 2$ is a key to judge the possible formation mechanisms. Unfortunately,
although hot gas is commonly observed in many (or most) local massive
elliptical galaxies \citep[see the review of][]{mathews03araa}, no study on
hot gas has been done for galaxies at $z\sim2$, limited by the sensitivity of
our current X-ray detectors.

\subsection{Metallicity Gradient}
\label{sub:metalgrad}

In the previous sections we have seen that, because of the age--metallicity
degeneracy, different assumptions for the metallicity gradient result in
different age gradients, given the observed color gradients. We have also seen
that uncertainties on the dust obscuration (both the total amount and its
gradient) do not affect the quantitative details of the relationship between
age and metallicity, given the current uncertainty. Thus, an interesting
question to ask is: given a realistic assumption for the metallicity gradient
at $z \sim 2$, namely one that is consistent with the metallicity gradient seen
in early--type galaxies $z \sim 0$ and with our current ideas on how
metallicity gradients evolve, what is the implied gradient of stellar
population age? What would such an age gradient tell us about the way 
$z\sim 2$ PEGs assembled and their subsequent
evolution, if they really are the progenitors of the local early--type
galaxies?

\subsubsection{Flat Metallicity Gradient}
\label{subsub:flat}

Under the assumption of a flat metallicity gradient, i.e. metallicity is
constant as a function of radius, the observed color gradients of the galaxies
imply a negative age gradient, namely the age of the stellar populations is
younger as the radial distance from the center increases, with average
gradient ${\rm \Delta log(t) / \Delta log(R) \sim -0.1}$. Stars located at
$\approx 10\times {\rm R_{eff}}$ from those in the center are, on average, $\sim 1$ Gyr
younger. The SSFR is also higher in the outer regions. For example, the
external rings in three of the galaxies, 19389, 22704 and 24626 have
SSFR$>{\rm 10^{-11}/yrs}$, larger than the global value we use to classify the
galaxies as passive. The younger stellar populations and larger specific star
formation rate in the outskirts of the galaxies could mean a later cessation
of star formation relative to the center, newer episodes of star formation or
accretion of younger stellar populations. 

It is interesting, at this purpose, to explore whether the residual star
formation in the outskirts can explain the apparent size evolution of massive
passively--evolving galaxies from $z\sim2$ to $z\sim0$, as discussed by recent
studies
\citep[e.g.,][]{daddi05,trujillo06,trujillo07,vandokkum08,cassata10}. To do
so, we simulate a galaxy with \sersic index $n=2.0$ and effective radius
${\rm R_{eff}=0.5}$ kpc (typical values for massive PEGs at $z\sim 2$, see by
\citet{cassata10}), central SSFR ${\rm 10^{-11} yr^{-1}}$ and the same SSFR
gradient as the one in Figure \ref{fig:covar}.  If this galaxy evolves from
$z=2$ to $z=0$ only through in-situ star formation, i.e. with no significant
accretion of external stars, our calculation shows that it cannot evolve
into today's typical massive early--type galaxies, which have $n=4$ and ${\rm
  R_{eff} \sim 2.5 kpc}$, since that would require the SSFR in the outskirts to be
$> 1.5$ dex higher than the central one SSFR, a much steeper gradient than our
observations seem to find. This implies that external mechanisms, such as
merger and accretion, are necessary to build the extended halos of massive
PEGs from $z\sim2$ to $z\sim0$. We also note that the assumption of flat
metallicity results in the steepest positive SSFR gradient compared to the
local metallicity gradient and the monolithic--collapse gradient that we will
discuss next. Thus, these two cases, too, would imply external mechanisms if
the $z \sim 2$ massive PEG are to evolve into the local early types.

If merger and accretion do drive the evolution, they must be
able to do so in a way that makes the flat metallicity gradient evolve into
the one observed in local early--type galaxies, ${\rm \Delta log(Z) / \Delta
  log(R) \sim -0.3}$, while at the same time cancel the negative age
gradient, since the age the stellar populations in local galaxies has 
very little but positive radial dependence \citep[e.g.,][]{tamura04,wuhong05,labarbera09}. While secular orbit mixing
could help explain today's flat age gradient, it seems hard to understand how
a flat metallicity gradient at $z\sim 2$ can evolve into the local one if 
major merger drives the evolution, since major merger is believed to 
flatten, not steepen the metallicity gradient \citep{kobayashi04}.

\subsubsection{Local Metallicity Gradient}
\label{subsub:local}
% \textit{Local metallicity gradient} 

If we assume that the $z \sim 2$ PEGs have the same metallicity gradient as their
local counterparts, the observed color gradients imply no age gradient,
as shown in the {\it middle} panel of Figure \ref{fig:covar}. In other words,
the radial dependence of metallicity and age of the the stellar populations of
the $z\sim 2$ galaxies is already similar to that of their local counterparts.
Furthermore, the implied gradient of SSFR is also flat. Thus, if merger or
accretion drive the evolution to $z\sim 0$, this must happen in a way
that maintains the gradients of metallicity and age roughly constant in
time. Since major merger appears to flatten the metallicity gradient
\citep{kobayashi04}, the assumption of the local metallicity gradient would
also imply a more gradual accretion process as the one responsible for the
apparent growth in size of PEG from $z\sim 2$ to the present
\citep[e.g.][]{vandokkum10}.

\subsubsection{Monolithic metallicity gradient} 
\label{subsub:monolithic}

The monolithic collapse is an idealized model in which a whole worth of stars
of a massive galaxy form during $\sim 1$ dynamical time scale. 
Although a recent monolithic collapse model by \citet{pipino10}
that allows certain scatter for the star formation efficiency would 
produce the metallicity gradient that agrees with the observation of 
local elliptical galaxies, earlier models by \citet{larson74} and 
\citet{carlberg84} define a maximum steepness boundary in the metallicity 
gradient slope--mass plane. 
% These galaxies are preferentially those with the highest SF efficiency at that given mass.
We discuss models by \citet{larson74} and 
\citet{carlberg84} here simply as the limiting case of a class of assembly 
mechanisms capable to produce the steepest metallicity gradient across 
the galaxy. 

During the monolithic
collapse, stars begin to form everywhere in the collapsing cloud and, once
formed, remain in their orbits with little net inward motion, while the gas
keeps sinking to the center of the galaxy due to dissipation. While getting
closer to the center, the gas become more and more enriched by the rapidly
evolving massive stars. Consequently, stars formed in the central regions are
more metal rich than those formed in the outskirts. Stellar feedback tends 
to reduce
the inflow of gas and hence reduce the metallicity gradient. But gas outflows
occur earlier and more effectively at large galactocentric distance than in
the center due to lower escape velocity, lowering the star--formation rate at
larger distance and contribute to create a strong negative metallicity
gradient and a positive age one.

Under the assumption of the monolithic collapse metallicity gradient, the
observed color gradients of our sample galaxies indeed imply a positive age
gradient such that the stars at ${\rm R\approx 10 \times R_{eff}}$ in are $\sim 0.5$ Gyr
older than those in the central regions. We could directly test this
prediction of the monolithic collapse (or equivalent scenarios), if we were
able to independently measure the age gradient, something that is not possible
with the present data. The monolithic collapse metallicity gradient also
implies a weak SSFR gradient for our galaxies such that the outer regions at
${\rm R\sim 10\times R_{eff}}$ have SSFR $\sim 0.5$ dex lower than in the center. This
is qualitative consistent with the general feature of the model that stellar
feedback is more effective at larger radii than at the center at reducing the
star formation activity\footnote{We note that according to
  \citet{martinelli98} constant star formation efficiency with galactocentric
  distance can also explain the observed metallicity gradient and the
  correlation between colors/metallicity and escape velocity in early--type
  galaxies.}

If the PEGs observed at $z\sim 2$ formed through
mechanisms similar to monolithic collapse, their subsequent evolution must
be such to significantly reduce the magnitude of their metallicity gradient
and to a minor extent the age gradient, since the monolithic--collapse
gradient is much steeper \citep{larson74,carlberg84} than that observed in
local ellipticals \citep[e.g.,][]{peletier90b,idiart03,tamura03}. Major merger
provides such a mechanism \citet{kobayashi04}, although the effectiveness of
minor merger or gas accretion and subsequent star formation in diminishing
the steep metallicity gradient is not known. 
% (YICHENG VERIFY THIS). 

The monolithic collapse model also predicts that the slope of the metallicity
gradient, and hence the color gradient, depends on the mass of the galaxies,
because a deeper potential well is more effective at retaining metals in the
center than a shallower one and thus make more metal--rich stars \citep[e.g.,][]{tortora10}.  As shown in
Figure \ref{fig:cg}, the color gradient of our galaxies does not show any
obvious dependence on the stellar mass, to the extent that this quantity is a
good proxy for the galaxies' total mass. The relative high scatter in our
small sample and the limited stellar mass dynamic range that it probes,
however, might hide such signal. We will return to the correlation of the
color gradient with the galaxies' properties using a much larger and
significantly deeper sample extracted from the new WFC3 CANDELS survey.
% (YICHENG, CAN YOU SPECIFY IN QUANTITATIVE TERMS THE EXPECTED CORRELATION
% BETWEEN METALLICTY GRADIENBT SLOPE AND STELLAR MASS?).

Independent measures of the age gradient of the stellar populations would also
test if mechanisms similar to the monolithic collapse play a role in
assembling the $z\sim 2$ PEG, since in this case the stars in the central
regions would be younger than those in the outskirts. In fact, such an
inverted age gradient is requested by our data in order to reproduce the
observed color gradients if the metallicity gradient of the monolithic
collapse is assumed, since this would result in significantly steeper color
gradients, as we have directly verified.

\subsection{Formation of Passively Evolving Galaxies at z$\sim$2}
\label{sub:pegformation}

It is more likely, however, that massive PEGs at z$\sim$2 are formed through
gas--rich major mergers rather than a single collapse process. Recently, 
\citet{wuyts10} analyzed SPH simulations of gas--rich mergers and their 
remnants that are treated with radiative transfer. They predicted that 
quiescent compact galaxies at z$\sim$2 should typically show red cores 
and their color gradients should be a superposition of age, dust, and 
metallicity gradients. They found that in the gas--rich merger scenario, 
stars in the galactic center are formed during final coalescence out of more 
obscured and enriched gas. The dust and metallicity gradients compensate the 
positive age gradient (young center and old outskirts) so that their cores 
are typical when these galaxies are classified as passive systems. They 
also predicted that the strength of the color gradient to be correlated with 
galaxy's integrated color. All these predictions agree well with our 
observations and serve as important evidence of the validity of gas--rich
major mergers.

Even if gas--rich major merger is considered as the formation mechanism of 
massive PEGs at z$\sim$2, their subsequent evolution is still ambiguous. 
\citet{kobayashi04} predicts that gas--rich merger can effectively flatten
the metallicity gradient to the one of local early--types. Therefore,
mechanisms that can significantly flatten the metallicity gradient, such as
major merger, are not required in the subsequent evolution. \citet{wuyts10}, 
however, find that the typical metallicity gradient in the simulated z~2 
gas-rich merger remnants is steeper than the typical metallicity gradient 
of local early--types. This requires major mergers in the subsequent evolution
to flatten the metallicity gradient, unless other mechanisms (accretion 
and minor merger) are proved to be capable to flatten the metallicity 
gradient too.

Overall, with the current data it is not possible to conclusively rule out or
validate any of the three cases of metallicity gradients and possible 
formation mechanisms of massive PEGs at z$\sim$2 that we have
discussed. Passively evolving galaxies appear to undergo substantial
structural evolution from $z\sim 2$ to $z\sim 0$ that reduces their
compactness and stellar density \citep[e.g.,][]{daddi05, trujillo06,
  trujillo07, vandokkum08, cassata10}
If major merging events are the driver of this evolution, to the extent
that we understand how merger rearranges gradients of metallicity and age, it
seems unlikely that the $z\sim 2$ PEG have flat metallicity gradients, since
subsequent merger can only keep it flatter, not steepen it to an extent
required to match the observed one in local ellipticals. Of course, the size
evolution can be driven by less dramatic minor merging events or continuous
accretion, as some have suggested \citep{vandokkum10}. We do not know what
these mechanisms would imply for the evolution of the metallicity and age
gradients compatible with the color gradients observed at $z\sim 2$ if they
have to evolve into those observed at $z\sim 0$. Finally, we remind that our
discussion is based on resolved photometry that only covers the UV/Optical rest
frame. Future high--resolution observations with JWST extending the wavelength
baseline to the near and mid--IR will allow us to considerably reduce the 
extent of the age--metallicity degeneracy, and help us constraint a
self--consistent evolutionary scenario for the assembly of the $z\sim 2$ PEG,
as well as their subsequent evolution.

\section{Summary}
\label{summary}

We have discussed the implications of the detection of color gradients in
early--type galaxies at $z\sim2$ from deep high--angular resolution images at
optical and near--IR wavelengths obtained with {\it HST} and the ACS and WFC3
cameras.

In particular, we have measured resolved rest--frame UV-optical colors of a
sample of six massive ($>{\rm 10^{10}M_\odot}$) and passively evolving
(SSFR$<{\rm 10^{-11} yr^{-1}}$) galaxies at $1.32<z<2.42$. After defining for
each galaxy a set of concentric apertures that optimally sample the observed
gradient of colors, we have carried out fits to spectral population synthesis
models using the available seven--band (BVizYJH) photometry to derive how dust
obscuration (E(B-V)), mean age, specific star formation rate (SSFR), and
stellar mass (${\rm M_{star}}$) vary with the galactocentric distance. We have
then used this information to discuss possible evolutionary scenarios for
these galaxies in light of recent results on the apparent evolution of their
morphological evolution and on theoretical expectations on how merger
modifies existing gradients of metallicity and stellar age. This paper can be
summarized as follows:

\begin{enumerate}

\item
Color gradients could be measured over scales that typically go up to $\approx
10\times {\rm R_{eff}}$, where ${\rm R_{eff}}$ is the effective radius of the \sersic
profile. The {\it HST} images show that the inner regions of these galaxies
have redder rest-frame UV--optical colors (U-V, U-B and B-V) than their outer
parts.

\item
The slopes of the color gradients have no dependence on the redshift and
stellar mass of the galaxies. However, they have a mild dependence on the
global dust extinction and rest-frame U-V color of the galaxies. Galaxies with
larger E(B-V) or redder U-V color tend to have steeper color gradients.

\item 
The slopes of the color gradients of these galaxies are generally steeper than
that of local early-type galaxies.

\item 
We investigate whether the variation of a single parameter (age, extinction,
or metallicity) along radius can be used to explain the observed color
gradients. Using the single stellar population model, we find that the
variation of any single parameter cannot simultaneously fit the three observed
color gradients (U-B, U-V and B-V) with the maximum likelihood. We conclude
that the observed color gradients of massive PEGs at $z\sim2$ cannot be
explained by a single gradient of age, extinction or metallicity and should be
originated from an interplay of gradients of the three parameters.

\item
The fits of spatially resolved stellar populations to the spectral population synthesis models are run under three
assumptions of metallicity gradients: (1) a flat metallicity gradient (${\rm
  \Delta log(Z) / \Delta log(R) = 0}$), (2) the metallicity gradient of local
early-type galaxies (${\rm \Delta log(Z) / \Delta log(R) = 0.25}$), and (3)
the gradient predicted by the monolithic collapse (${\rm \Delta log(Z) /
  \Delta log(R) = 0.5}$).

\item 
Regardless of the assumptions on metallicity, a modest gradient of dust
obscuration is always implied from the fits in the sense that the central
regions of the galaxies have slightly higher dust obscuration than the outer
parts, with an average gradient of ${\rm \Delta E(B-V) / \Delta log(R) \sim
  -0.07}$, if the starburst obscuration law by \citet{calzetti94, calzetti00} 
is used. Other
extinction laws that we have tested (MW, SMC, LMC) result in smaller
obscuration gradients. Overall, both the absolute value of dust obscuration
and its gradient are small, however, consistently with the present--day
early--type galaxies, where dust generally has small, if any effects on the
observed colors. It appears that once a galaxy has become passive, for
whatever physical mechanisms, dust obscuration ceases to play a significant
role in the determining the UV/Optical SED.

\item 
While dust obscuration contributes in small measure to the observed color
gradients of the $z\sim 2$ galaxies, its presence does not seem to affect the
general age--metallicity degeneracy in the sense that the implied gradient of
age derived from a given assumption for the gradient of metallicity does not
depend on how dust is treated, i.e. if forced to a fixed value or left as a
free parameter in each annulus, or on the adopted extinction law. Whatever
inference on the age or on the metallicity gradient is made, after assuming
one or the other parameter, does not seem to appreciably depend on the
assumption on dust obscuration.

\item
Due to the age--metallicity degeneracy, the derived age gradients are strongly
coupled with the assumed metallicity gradients: (1) assuming a flat
metallicity gradient, the outer regions of the galaxies are younger than the
inner regions with a age gradient of ${\rm \Delta log(t) / \Delta log(R) \sim
  -0.1}$; (2) assuming the metallicity gradient observed in local early--type
galaxies, the stellar populations in the outer regions have same age as those
in the inner regions; and (3) for the metallicity gradients predicted by the
monolithic collapse, the outer regions are older than the inner regions, with
the average age gradient ${\rm \Delta log(t) / \Delta log(R) \sim
  0.15}$. Their specific star--formation rate is also $\sim$0.5 dex lower than
that in the inner regions.

\item 
The mass--size (or equivalently mass--stellar density) relationship of the
$z\sim 2$ galaxies cannot evolve into the local one only through
in--situ star formation driven by the small observed star--formation activity
(SSFR$<10^{-11}$ yr$^{-1}$ or less). This implies the accretion of stellar
mass from outside \citep{vandokkum10}. 

\item
Overall, with the current data it is not possible to conclusively rule out or
validate any of the three cases of metallicity gradients that we have
considered. A major source of uncertainty is the fact that major merger
rearranges the gradients of metallicity and age on a short time scale, while
less dramatic events such as minor merger or a more continuous accretion
might induce a more ``secular'' evolution of these properties. Passively
evolving galaxies appear to undergo substantial structural evolution from
$z\sim 2$ to $z\sim 0$ that reduces their compactness by a factor of 3--5 and
their stellar density by $\sim 2$ orders of magnitude 
\citep[e.g.,][]{daddi05, trujillo06, trujillo07, vandokkum08, cassata10}
% (YICHENG PUT HERE RELEVANT REFERENCES). 
If major merging events are the driver of this
evolution, then, to the extent that we understand how merer rearranges
gradients of metallicity and age, it seems unlikely that the $z\sim 2$ PEG
have flat metallicity gradients, since subsequent merger can only keep it
flatter, not steepen it to an extent required to match the observed one in
local ellipticals. Of course, the size evolution can be driven by minor
merger/accretion, as some have suggested \citep{vandokkum10}. In this case,
we have much less guidance in inferring which metallicity and age gradients
are compatible with the color gradients observed at $z\sim 2$ if they have to
evolve into those observed at $z\sim 0$.

\item
While it is possible that the subsequent evolution reconciles the metallicity
and age gradient emerging from the monolithic collapse to those observed at
$z\sim 0$, the observations do not seem to show any correlation between the
strength of the color gradient and the stellar mass, which is predicted if the
$z\sim 2$ PEG formed through such a mechanism. The inherent statistical noise
in a sample as small as ours, and the fact that the sample itself only covers
a small dynamic range in mass, can very well hide any such correlation. We do
observe, however, a correlation between the color gradient and the dust
obscuration (E(B-V)), even if such parameter is generally much less accurately
estimated with broad--band SED fitting than the stellar mass, which is the
most accurate one. This seems to support the lack of a correlation between the
color gradient and the stellar mass, and thus argue against the monolithic
collapse, or any formation mechanism capable to produce an equally steep
metallicity gradient, as responsible for the formation of the $z\sim 2$
PEG. We will return on this subject using substantially large samples of such
sources from the CANDELS project.

\item
The metallicity gradient of the galaxies could be either close to that of the
local early--type galaxies or flat. In the first case, the subsequent
evolution must be such to preserve the metallicity gradient, which would seem
to rule out major merger. In the second case, the evolution must create the
gradient. This also seems to rule out major merger, since it can only
flatten, not steepen, the gradient. 

\end{enumerate}

We thank Houjun Mo, Eric Gawiser and Massimo Stiavelli for useful comments and 
discussions. YG, MG, PC and SS acknowledge support from NASA grants 
HST-GO-9425.36-A, HST-GO-9822.45-A, and HST-GO-10189.15-A, awarded by the 
Space Telescope Science Institute, which is operated by the Association of 
Universities for Research in Astronomy, Inc. (AURA) under NASA contract 
NAS 5-26555.

%\bibliographystyle{apj}
%\bibliography{references}

\end{document}